\documentclass[12pt]{article}
\pdfoutput=1
\usepackage{titling}
\usepackage{amsmath}
\usepackage{slashed}
\usepackage{amssymb}
\usepackage{epsfig}
\usepackage{graphicx}
\usepackage{multirow}
\usepackage{color}
\usepackage[normal]{subfigure}
\usepackage{rotating}
\usepackage{hyperref}
\usepackage[margin=0.9in]{geometry}
\usepackage[table]{xcolor}
\usepackage{enumitem}
\usepackage[utf8x]{inputenc}
\usepackage[compress,numbers,sort]{natbib}
\usepackage{authblk}
\usepackage{colortbl}
\usepackage{pdflscape}
\usepackage{color}
\usepackage{mathtools}
\usepackage{tabu}
\usepackage{braket}

\def\eq#1{{Eq.~(\ref{#1})}}
\def\eqs#1#2{{Eqs.~(\ref{#1})--(\ref{#2})}}
\def\fig#1{{Fig.~\ref{#1}}}

\def\Table#1{{Table~\ref{#1}}}
\def\Tables#1#2{{Tables~\ref{#1}--\ref{#2}}}

\def\Re{\mbox{Re}}

\renewcommand{\bar}{\overline}
\def\q3l{q_{3L}}
\def\bq3l{\overline{q}_{3L}}
\def\l3l{\ell_{3L}}
\def\bl3l{\overline{\ell}_{3L}}
\def\blep[#1]{\overline{L}_{#1}}
\def\lep[#1]{L^{#1}}
\def\bql[#1]{\overline{Q}_{#1}}
\def\ql[#1]{Q^{#1}}
\def\vld[#1]{V^{\dagger}_{L #1}}
\def\vl[#1]{V^{#1}_{L}}
\def\vqd[#1]{V^{\dagger}_{Q #1}}
\def\lr[#1]{#1_{R}}
\def\blr[#1]{\overline{#1}_{R}}
\def\cq#1{C^{q}_{#1}}
\def\cqs#1{C^{q}_{S#1}}
\def\cqr#1{C^{q}_{R#1}}
\def\cl#1#2{C^{\ell}_{#1#2}}

\def\clr#1{C^{\ell}_{R#1}}
\def\clt#1{C^{\ell}_{T#1}}
\def\cqq#1{C^{qq}_{#1}}
\def\ckm#1{V_{#1}}
\def\ckmd#1{V^{*}_{#1}}
\def\vq{V_{Q}}

\def\vl{V_{L}}

\def\epsl#1{\epsilon^{\ell}_{#1}}
\def\epsq#1{\epsilon^{q}_{#1}}
\def\epsqp{\epsilon_{q}'}
\def\epslp{\epsilon_{\ell}'}
\newcommand{\be} {\begin{equation}}
\newcommand{\ee} {\end{equation}}
\newcommand{\ba} {\begin{eqnarray}}
\newcommand{\ea} {\end{eqnarray}}

\newcommand{\cA} {\mathcal A}
\newcommand{\cB} {\mathcal B}
\newcommand{\RK} {R^{\mu/e}_K}

\newcommand{\published}[1]{%
\gdef\puB{#1}}
\newcommand{\puB}{}

\title{\Large\bf  Semi-leptonic $B$-physics anomalies: \\[0.2 cm]
a general EFT analysis within $U(2)^n$  flavor symmetry }

\author[1]{\large Marzia Bordone\thanks{marzia.bordone@physik.uzh.ch}}
\author[1]{\large Gino Isidori\thanks{isidori@physik.uzh.ch}}
\author[1]{\large Sokratis Trifinopoulos\thanks{trifinos@student.ethz.ch}}
\affil[1]{\emph{\normalsize Physik-Institut, Universit\"at Z\"urich, CH-8057 Z\"urich, Switzerland}}

\date{}

\published{
\begin{flushright}
ZU-TH 02/17  \\
February 2017
\end{flushright}
\vskip 2cm }

\begin{document} 

\maketitle

\begin{abstract}
\noindent
We analyse the recent hints of Lepton Flavor Universality violations 
in semi-leptonic $B$ decays within a general EFT based on 
a $U(2)^n$ flavor symmetry acting on the light generations of SM fermions.
We analyse in particular the consistency of these anomalies with 
the tight constraints on various low-energy observables in $B$ and $\tau$
physics. We show that, with a moderate fine-tuning, 
a consistent picture for all low-energy observables 
can be obtained under the additional dynamical assumption 
that the NP sector is coupled preferentially to third generation SM fermions.
We discuss how this dynamical assumption can be implemented in general 
terms within the EFT, and we identify a series of observables 
in $\tau$ decays which could provide further evidences 
of this NP framework. 
\end{abstract}

\clearpage
\tableofcontents
\clearpage

\section{Introduction}
\label{sec:Introduction}
The hints of Lepton Flavor Universality (LFU) violations in semi-leptonic $B$ decays
are among the most interesting and persisting deviations from the Standard Model (SM)
reported by experiments in the last few years. 
The statistically most significant results are encoded by the following three ratios:
\ba
R^{\tau/\ell}_{D^*} &=&\frac{ \cB(B \to D^* \tau \overline{\nu})_{\rm exp}/\cB(B \to D^* \tau \overline{\nu})_{\rm SM} }{ \cB(B \to D^* \ell \overline{\nu} )_{\rm exp}/ \cB(B \to D^* \ell \overline{\nu} )_{\rm SM} } = 
 1.23 \pm 0.07~, \label{eq:RDexp}  \\
R^{\tau/\ell}_{D } &=& \frac{ \cB(B \to D  \tau \overline{\nu})_{\rm exp}/\cB(B \to D  \tau \overline{\nu})_{\rm SM} }{ \cB(B \to D  \ell \overline{\nu} )_{\rm exp}/ \cB(B \to D  \ell \overline{\nu} )_{\rm SM} } =   1.34 \pm 0.17~, \label{eq:RDSexp}  \\
\RK &=&  \left. \frac{ \cB(B \to K \mu\overline{\mu})_{\rm exp} }{ \cB(B \to K e \bar{e} )_{\rm exp} } \right|_{q^2\in[1,6]{\rm GeV}} =  0. 745 {}^{+0.090}_{-0.074} \pm 0.036~,
\label{eq:RKexp}
\ea
where $\ell$ generically denotes a light lepton ($\ell=e, \mu$).\footnote{~The first two results follow for the 
HFAG averages~\cite{Amhis:2016xyh} of Babar~\cite{Lees:2013uzd}, Belle~\cite{Hirose:2016wfn}, and LHCb data~\cite{Aaij:2015yra},
namely $\cB(B \to D^* \tau \overline{\nu})/\cB(B \to D^* \ell \overline{\nu} )_{\rm exp} =  0.310 \pm 0.017$  
and $\cB(B \to D  \tau \overline{\nu})/\cB(B \to D  \ell \overline{\nu} )_{\rm exp} = 0.403 \pm 0.047$,  
together with the corresponding theory predictions, $\cB(B \to D^* \tau \overline{\nu})/\cB(B \to D^* \ell \overline{\nu} )_{\rm SM} = 0.252\pm 0.003$~\cite{Fajfer:2012vx}
and  $\cB(B \to D  \tau \overline{\nu})/\cB(B \to D \ell \overline{\nu} )_{\rm SM} =  0.300\pm 0.008$~\cite{Aoki:2016frl}.
The latter result, based on LHCb data only~\cite{Aaij:2014ora}, should be compared with 
SM expectation $\RK =1.00 \pm 0.01$~\cite{Bordone:2016gaq}.}

In addition to these LFU ratios, whose deviation from unity would clearly signal physics beyond the  
SM, semi-leptonic $B$ decay data exhibit other tensions with the SM predictions. 
Most notably, a deviation of about $3\sigma$ has been reported 
 by LHCb~\cite{Aaij:2015oid} on the so-called $P_5^\prime (B\to K^*\mu\overline{\mu})$
differential observable.
This result is also compatible with recent Belle data~\cite{Wehle:2016yoi},
although the latter have a smaller statistical significance. The $P_5^\prime$  anomaly alone 
is not an unambiguous signal of new physics, given the non-negligible uncertainties 
affecting its SM prediction~\cite{Ciuchini:2015qxb}. However, it is interesting to note that all available 
$b\to s \ell\overline{\ell}$ data (including the ratio $\RK$ reported above)
turn out to be in better agreement with the corresponding theory predictions under the assumption 
of a single lepton-flavor non-universal short-distance amplitude 
affecting only the muonic modes (for an updated discussion see 
e.g.~Ref.~\cite{Descotes-Genon:2015uva,Altmannshofer:2015sma,Hurth:2016fbr} 
and references therein).

These deviations from the SM have triggered a series of theoretical speculations
about possible New Physics (NP) interpretations.  In particular, attempts to provide 
a combined/coherent explanation for  both charged- and neutral-current anomalies have been
presented  in Ref.~\cite{Bhattacharya:2014wla, Alonso:2015sja, Greljo:2015mma, Calibbi:2015kma,Bauer:2015knc,
Fajfer:2015ycq, Barbieri:2015yvd, Das:2016vkr, 
Boucenna:2016qad,Becirevic:2016yqi,  Hiller:2016kry, Bhattacharya:2016mcc}.
Among them, a particularly interesting class is that of models 
based on a $U(2)^n$ flavor symmetry, acting on the light generations of SM fermions~\cite{Barbieri:2011ci,Barbieri:2012uh},
and new massive vector mediators around the TeV scale 
(either colorless $SU(2)_L$ triplets~\cite{Greljo:2015mma},
 or  $SU(2)_L$ doublet leptoquarks~\cite{Barbieri:2015yvd}).
Beside providing a good description of low-energy data, 
these mediators could find a consistent UV completion in the context of 
strongly-interacting theories with new degrees of freedom at the TeV 
scale~\cite{Buttazzo:2016kid,Barbieri:2016las}.  
 
While these NP interpretations are quite interesting, their compatibility with 
high-$p_T$  data from the LHC  and other precision low-energy observables
 is not trivial. On the one hand, it has been pointed out that 
high-$p_T$ searches of  resonances decaying into a $\tau\overline{\tau}$  pair  ($pp \to\tau\overline{\tau} + X$) 
represent  a very stringent constraint for virtually any 
model addressing the $R^{\tau/\ell}_{D^(*)}$ anomalies~\cite{Faroughy:2016osc}.
On the other hand, the  consistency with LFU tests and 
the bounds on Lepton Flavor Violation  (LFV) from $\tau$ decays, 
after taking into account quantum corrections, seems to be problematic~\cite{Feruglio:2016gvd}.
Last but not least, in all the explicit models constructed so far, a non-negligible 
amount of  fine-tuning seems to be unavoidable in order 
to satisfy the constraints from $B_s$ and $B_d$ meson-antimeson 
mixing (see, in particular, Ref.~\cite{Barbieri:2015yvd,Buttazzo:2016kid}).
 
The compatibility with collider searches is certainly a serious issue; however, 
it should not be over-emphasized especially in the context of strongly interacting 
theories, where the extrapolation from low-energy data to the on-shell production 
of the new states is subject to sizable uncertainties.  On the contrary, the 
compatibility of these anomalies with other low-energy data is a question that 
can be addressed in a model-independent way using an appropriate 
Effective Field Theory (EFT) approach.
The purpose of this paper is to revisit the consistency and the compatibility 
of the anomalies reported in Eqs.~(\ref{eq:RDexp})--(\ref{eq:RKexp})
with other low-energy data,  employing a general EFT approach based 
on the  $U(2)^n$ flavor symmetry. 

As it appeared clear from the first $U(2)^n$ based 
analyses~\cite{Greljo:2015mma,Barbieri:2015yvd}, the flavor symmetry alone 
is not enough to guarantee  a natural explanation of   $B$-physics anomalies
in a general EFT approach. 
Additional dynamical assumptions are needed to explained the observed 
hierarchy among the various effective operators. Our goal is to discuss in general terms 
possible power-counting schemes to justify these  hierarchies 
and, within such schemes, to quantify the amount of fine-tuning necessary 
to obtain a satisfactory description of all low-energy data.

The paper is organised as follows. In Section~\ref{sect:setup} we define the low-energy  EFT 
and provide a complete list of the four-fermion operators with the 
inclusion of at most one lepton spurion and one (or two) quark spurion(s) contributing to $\Delta F=1$ (or $\Delta F=2$) processes.
The bounds on these operators from the relevant low-energy observables are discussed in Section~\ref{sect:obs}.
In Section~\ref{sect:EFTexp} we analyse these bounds and determine a consistent power-counting scheme that allow us 
justify the observed hierarchies. 
In Section~\ref{sect:HO} we discuss selected observables receiving leading contributions from operators with 
two lepton spurions (among which $\RK$), further testing the consistency of the proposed power-counting 
scheme. The final results, with a quantification of the fine-tuning needed to reconcile anomalies and bounds,
are summarised in the Conclusions. 

\section{Setup}
\label{sect:setup}
The EFT we are considering is characterised by the SM field content,
the  SM gauge symmetry ($SU(3)_c\times SU(2)_L\times U(1)_Y$), and a global flavor symmetry ${\mathcal G}_{\rm flavor}$, 
that we can decompose as follows
\be
{\mathcal G}_{\rm flavor} = U(2)_q \times U(2)_\ell \times {\mathcal G}_{R}~.
\label{eq:flavsymm}
\ee
The left-handed SM fermions ($q^i_L$ and $\ell^i_L$)  are singlets under ${\mathcal G}_{R}$
and have the following transformation properties under $U(2)_q \times U(2)_\ell$:
\begin{eqnarray}
&&   Q \equiv (q^1_L, q^2_L) \sim (2,1)~,  \qquad  q_{3L}  \equiv q^3_L \sim(1,1)~, \\ 
&&    L  \equiv (\ell^1_L, \ell^2_L) \sim (1,2)~,  \qquad  \ell_{3L} \equiv  \ell^3_L \sim (1,1)~.
\end{eqnarray}
The third-generation  right-handed fermions ($t_R$, $b_R$ and $\tau_R$)
are all singlets of the complete group ${\mathcal G}_{\rm flavor}$. 
Various options are possible as far as the action of $ {\mathcal G}_{\rm flavor}$
on the right-handed light-generation fermions is concerned. The simplest choice 
is the MFV-like~\cite{DAmbrosio:2002vsn} setting  ${\mathcal G}_{R} = U(2)_{u_R}  \times U(2)_{d_R}   \times U(2)_{e_R}$,
such that $E = (\mu_R, e_R)$ transforms as a doublet of $U(2)_{e_R}$, 
and similarly for right-handed light quarks.
But other options, where $\mu_R$ and $e_R$ belong to the same non-trivial representation of 
a non-Abelian  subgroup, leads to equivalent results. 

We further consider two breaking spurions of the flavor symmetry, $\vq$ and $\vl$, 
transforming, respectively,  as (2,1) and (1,2) of $U(2)_q \times U(2)_\ell$. 
The structure of $\vq$ can be connected to the CKM matrix ($V$) up to an overall normalization
factor~\cite{Barbieri:2011ci}:
\be 
\vq \equiv (V_{Q_1}, V_{Q_2}) = |V_Q | \times \left( \frac{V^*_{td}}{V^*_{ts}} , 1   \right), 
\ee
with $|V_Q |$ expected to be of $O(|V_{ts}|)$. 
In the case of $\vl$, in the absence of a clear connection to the entries in the lepton Yukawa couplings,
and given the strong universality bounds in processes involving electrons,
we assume the following hierarchical structure:
\be
\vl \equiv (V_{L_1}, V_{L_2}) = |V_L | \times \left( 0, 1 \right)~,
\ee
with $|V_L | \ll 1$ (an estimate of the  maximal allowed value 
for $|V_{L_1}/V_{L_2}|$ is presented in Section~\ref{sect:Ve_max}).

So far we have not specified the flavor basis of the left-handed fermion doublets, 
or better how we define the $U(2)_q \times U(2)_\ell$ singlets. In the lepton case, 
the natural choice is provided by the charged-lepton mass-eigenstate basis 
(or by identifying $\tau_L$ as the $U(2)_\ell$ singlet). In the quark sector the situation is more ambiguous.
In principle, any linear combination between down- and up-quark mass eigenstates is equally valid.
For the sake of simplicity, we assume as reference basis the down-quark mass-eigenstate basis.
This corresponds to identifying  the $U(2)_q$ singlet and doublet as\footnote{~In Eq.~(\ref{eq:basis})
we write explicitly the two electroweak doublet components of the doublets.}
\be
q_{3L} = \left( \begin{array}{c} V^*_{k b} u^k_L \\  b_L \end{array} \right)  \qquad {\rm and} \qquad
Q^i  = \left( \begin{array}{c} V^*_{k i} u^k_L \\  d^i_L \end{array} \right)~, \quad
  i= \{1,2 \} \equiv \{d,s \}~.
  \label{eq:basis}
\ee
A ``natural" change of basis is equivalent to the following shift in $q_{3L}$ 
\be
q_{3L}  \to q^\prime_{3L}  = \cos(\theta) q_{3L} + \sin(\theta) \vqd[i] Q^i ~,
\ee
where $\theta$ is an arbitrary angle.   
As a result, we can consider natural (non fine-tuned) the EFT constructions 
if operators without spurions and corresponding terms obtained with the replacement 
$q_{3L} \to \vqd[i] Q^i$ have coefficients of similar size.

\begin{table}[t]
\centering
\begin{tabu}{c|[1.5pt]c|[1.5pt]l}
&Operator                                                                                                                         & Relevant low-energy observables
                                                 \\ \tabucline[1.5pt]{-}
$\mathcal{O}^{qq}_{01}$ &  $\left[\bq3l\gamma^{\mu}\ql[i]\vqd[i]\right]^2$  &  $\Delta M_{B_d}$,   $\Delta M_{B_s}$  \\ \hline
$\mathcal{O}^{qq}_{02}$& $\left[\bq3l\sigma^{a}\gamma^{\mu}\ql[i]\vqd[i]\right]^2$ &  $\Delta M_{B_d}$,   $\Delta M_{B_s}$  \\ 
\tabucline[1.5pt]{-}
\end{tabu}
\caption{Four-quark operators contributing to $\Delta F=2$ amplitudes with  at most two quark spurions.}
\label{tbl:quark mixing operators}
\end{table}

\subsection{The basis of  effective operators}
\label{sect:basis}

In addition to the symmetries discussed above, we impose the conservation of baryon and lepton number, 
and we consider higher-dimensional operators up to dimension six.  
The EFT we are considering can thus be written as 
\be
\mathcal{L}_{\rm EFT} =  \mathcal{L}_{\rm SM}  + \frac{4 G_F}{\sqrt{2}}  \sum_i C_i \mathcal{O}_{i}~,
\label{eq:LEFT}
\end{equation}
using the Fermi scale, $v_F = (4 G_F/\sqrt{2})^{-1/2} \approx 174$~GeV, as overall dimensional normalization factor. 
With such choice we reabsorb the value of the EFT effective scale ($\Lambda$) inside the Wilson coefficients, 
whose  natural size in absence of specific suppression factors is $O(v_F^2/\Lambda^2)$.

The effective operators $\mathcal{O}_{i}$ can be separated into three main categories: i) operators with no fermion fields;
ii) operators with two fermion fields (plus Higgs or gauge fields); iii) four-fermion operators. The first two categories 
contain a small number of operators and are not particularly interesting to the processes we are 
considering.\footnote{~As discussed in the introduction, we focus our attention only low-energy processes. 
We do not include in this category precision electroweak tests at the $Z$-pole, which would be sensitive to four-fermion
operators at the one-loop level~\cite{Feruglio:2016gvd}, but also to operators of the first two categories.}
Within the class of four-fermion operators we can identify four interesting sub-categories, whose lists of operators, with the 
inclusion of at most one lepton spurion and one quark spurion (or two quark spurions in the case of $\Delta F=2$ operators), 
are  reported in Table~\ref{tbl:quark mixing operators}--\ref{tbl:leptonic operators}. 
For each operator we indicate the low-energy  processes that can provide the most stringent constraint.

In the case of semi-leptonic operators  we do not  list explicitly those with a pair of right-handed light quarks since  they do not
give rise to signatures different from those of the operators already listed and, in addition, 
can be assumed to be suppressed under natural dynamical assumptions. For similar reasons, despite 
we have explicitly  listed in Table~\ref{table:sl_2}--~\ref{tbl:leptonic operators} tensor operators, we will
ignore their effects in the phenomenological analysis of $b\to c \tau \bar\nu$ and $b\to s \tau \bar\tau$ transitions. 

\begin{table}[t]
\centering
 {\def\arraystretch{1.2}
\begin{tabu}{c|[1.5pt]c|[1.5pt]l}
&Operator                                                                                                                         & Relevant low-energy processes                                         \\ \tabucline[1.5pt]{-}
$\mathcal{O}^{q}_{01}$ &  $\left(\bq3l\gamma^{\mu}\q3l\right)\left(\bl3l\gamma_{\mu}\l3l\right)$  & --- ($\nu_\tau N \to \nu_\tau N$,  
$\Upsilon \to \tau\overline{\tau}$)   \\ \hline
$\mathcal{O}^{q}_{02}$& $\left(\bq3l\sigma^{a}\gamma^{\mu}\q3l\right)\left(\bl3l\sigma_{a}\gamma_{\mu}\l3l\right)$ &  $b\to c\tau\overline{\nu}$  \\ \hline 
$\mathcal{O}^{q}_{03}$ & $\left(\bq3l\gamma^{\mu}\q3l\right)\left(\blep[i]\gamma_{\mu}\lep[i]\right)$ & ---  ($\nu_{\ell} N \to \nu_{\ell} N$,  
$\Upsilon \to \ell\overline{\ell}$)  \\ \hline
$\mathcal{O}^{q}_{04}$ & $\left(\bq3l\sigma^{a}\gamma^{\mu}\q3l\right)\left(\blep[i]\sigma_{a}\gamma_{\mu}\lep[i]\right)$ &    $b\to c\mu\overline{\nu}$ \\ \hline
$\mathcal{O}^{q}_{05}$ & $\left(\bql[i]\gamma^{\mu}\ql[i]\right)\left(\bl3l\gamma_{\mu}\l3l\right)$ & ---   ($\nu_\tau N \to \nu_\tau N$,  
$\phi \to \tau\overline{\tau}$)  \\ \hline
$\mathcal{O}^{q}_{06}$ & $\left(\bql[i]\sigma^{a}\gamma^{\mu}\ql[i]\right)\left(\bl3l\sigma_{a}\gamma_{\mu}\l3l\right)$ & $\tau\to K\nu$,
$D \to \tau \nu$   \\ \hline
$\mathcal{O}^{q}_{07}$ & $\left(\bql[i]\gamma^{\mu}\ql[i]\right)\left(\blep[i]\gamma_{\mu}\lep[i]\right)$ & ---   ($\nu_\ell N \to \nu_\ell N$,  
$\phi \to \ell\overline{\ell}$) \\ \hline
$\mathcal{O}^{q}_{08}$ & $\left(\bql[i]\sigma^{a}\gamma^{\mu}\ql[i]\right)\left(\blep[i]\sigma_{a}\gamma_{\mu}\lep[i]\right)$ &  $ K\to  \ell\overline{\nu}$, 
$ K\to \pi \ell\overline{\nu}$,  $ \pi \to \ell\overline{\nu}$    \\ \hline
\tabucline[1.5pt]{-}
$\mathcal{O}^{q}_{11}$ & $\left(\bq3l\gamma^{\mu}\ql[i]\vqd[i]\right)\left(\bl3l\gamma_{\mu}\l3l\right)$       & $b\to s\tau\overline{\tau}$, $b\to s\nu\overline{\nu}$\\ \hline
$\mathcal{O}^{q}_{12}$ & $\left(\bq3l\sigma^{a}\gamma^{\mu}\ql[i]\vqd[i]\right)\left(\bl3l\sigma_{a}\gamma_{\mu}\l3l\right)$   & $b\to c\tau\overline{\nu}$, $b\to s\tau\overline{\tau}$, $b\to s\nu\overline{\nu}$, $\tau\to K\nu$\\ \hline
$\mathcal{O}^{q}_{13}$ & $\left(\bq3l\gamma^{\mu}\ql[i]\vqd[i]\right)\left(\blep[i]\gamma_{\mu}\lep[i]\right)$ & $b\to s\ell\overline{\ell}$, $b\to s\nu\overline{\nu}$\\ \hline
$\mathcal{O}^{q}_{14}$ & $\left(\bq3l\sigma^{a}\gamma^{\mu}\ql[i]\vqd[i]\right)\left(\blep[i]\sigma_{a}\gamma_{\mu}\lep[i]\right)$  & $b\to s\ell\overline{\ell}$, $b\to s\nu\overline{\nu}$     \\
\tabucline[1.5pt]{-}
$\mathcal{O}^{q}_{21}$ &  $\left(\bq3l\gamma^{\mu}\q3l\right)\left(\bl3l\gamma_{\mu}\lep[i]\vld[i]\right)$  & 
   $\Upsilon \to \tau\overline{\mu}$,  $\eta_{b}\to\tau\mu$
 \\ \hline
$\mathcal{O}^{q}_{22}$& $\left(\bq3l\sigma^{a}\gamma^{\mu}\q3l\right)\left(\bl3l\sigma_{a}\gamma_{\mu}\lep[i]\vld[i]\right)$ & 
  $\Upsilon \to \tau\overline{\mu}$, $\eta_{b}\to\tau\mu$  \\         \hline    
$\mathcal{O}^{q}_{23}$ &  $\left(\bql[i]\gamma^{\mu}\ql[i]\right)\left(\bl3l\gamma_{\mu}\lep[i]\vld[i]\right)$  & 
$\tau \to   \mu \rho  $,  $\tau \to   \mu \omega  $  \\ \hline
$\mathcal{O}^{q}_{24}$& $\left(\bql[i]\sigma^{a}\gamma^{\mu}\ql[i]\right)\left(\bl3l\sigma_{a}\gamma_{\mu}\lep[i]\vld[i]\right)$ & 
$\tau \to   \mu \rho  $,  $\tau \to   \mu \omega  $   \\     
\tabucline[1.5pt]{-}
 $\mathcal{O}^{q}_{31}$ &  $\left(\bq3l\gamma^{\mu}\ql[i]\vqd[i]\right)\left(\bl3l\gamma_{\mu}\lep[i]\vld[i]\right)$  & 
 $B_s \to \tau\overline{\mu}$ \\ \hline
$\mathcal{O}^{q}_{32}$& $\left(\bq3l\sigma^{a}\gamma^{\mu}\ql[i]\vqd[i]\right)\left(\bl3l\sigma_{a}\gamma_{\mu}\lep[i]\vld[i]\right)$ &
 $B_s \to \tau\overline{\mu}$    \\  
\tabucline[1.5pt]{-}
\end{tabu}}
\caption{Semi-leptonic four-fermion operators, with only left-handed currents and at most one lepton and/or one quark spurion.
The processes listed between brackets do not give appreciable bounds and are reported only for completeness.}
\label{table:sl_1}
\end{table}
 
In principle, the various effective operators mix under quantum corrections.
However,  as indicated in Eq.~(\ref{eq:LEFT}), we assume a rather low effective scale 
such that  no large  logarithms  are involved in the renormalization-group (RG) evolution.
This implies that in most cases these mixing effects can be neglected. 
The only exception are cases where an operator with a large coefficient 
(in particular those contributing to $R_{D^{(*)}}$)
mixes into a strongly constrained one (such as those contributing to leptonic $\tau$ decays),
as pointed out first in Ref.~\cite{Feruglio:2016gvd}.
Since no large logarithms are involved, 
we take into account this effects directly at the matrix-element level  
(i.e.~taking into account also one-loop matrix elements, when necessary).

\begin{table}[t]
\centering
 {\def\arraystretch{1.2}
\begin{tabu}{c|[1.5pt]c|[1.5pt]l}
&Operator                                                                                                                         &    Relevant low-energy processes                                               \\ \tabucline[1.5pt]{-}
 $\mathcal{O}^{q}_{R1}$ &  $\left(\bq3l\gamma^{\mu}\q3l\right)\left(\blr[\tau]\gamma_{\mu}\lr[\tau]\right)$  & 
 --- ($\Upsilon \to \tau\overline{\tau}$)     \\ \hline
  $\mathcal{O}^{q}_{R2}$ &  $\left(\bq3l\gamma^{\mu}\ql[i]\vqd[i]\right)\left(\blr[\tau]\gamma_{\mu}\lr[\tau]\right)$  & $b\to s\tau\overline{\tau}$  \\ \hline
    $\mathcal{O}^{q}_{R3}$ &  $\left(\bql[i]\gamma^{\mu}\ql[i]\right)\left(\blr[\tau]\gamma_{\mu}\lr[\tau]\right)$  & 
    --- ($ \tau N \to \tau N$)  \\  \hline
     $\mathcal{O}^{q}_{R4}$ &  $\left(\bq3l\gamma^{\mu}\q3l\right)\left(\overline{E}_j\gamma_{\mu}{E^j}\right)$  & 
     --- ($\Upsilon \to \ell\overline{\ell}$) \\ \hline
  $\mathcal{O}^{q}_{R5}$ &  $\left(\bq3l\gamma^{\mu}\ql[i]\vqd[i]\right)\left(\overline{E}_j\gamma_{\mu}{E^j}\right)$  & $b\to s\ell\overline{\ell}$  \\ \hline
    $\mathcal{O}^{q}_{R6}$ &  $\left(\bql[i]\gamma^{\mu}\ql[i]\right)\left(\overline{E}_j\gamma_{\mu}{E^j}\right)$  & 
    --- ($\phi \to \ell\overline{\ell}$)  \\ 
\tabucline[1.5pt]{-}
 $\mathcal{O}^{q}_{S1}$ &  $\left(\bl3l\lr[\tau]\right)\left(\blr[b]\q3l\right)$  &  $b\to c\tau\overline{\nu}$ \\ \hline
  $\mathcal{O}^{q}_{S2}$ &  $\left(\bl3l\lr[\tau]\right)\left(\blr[b]\ql[i]\vqd[i]\right)$  & $b\to c\tau\overline{\nu}$, $b\to s\tau\overline{\tau}$ \\ \hline
 $\mathcal{O}^{q}_{S3}$ &  $\left(\blep[i]\vl^i \lr[\tau]\right)\left(\blr[b]\q3l\right)$  &   $\eta_b \to \tau\overline{\mu}$     \\ 
\tabucline[1.5pt]{-}
 $\mathcal{O}^{q}_{T1}$ &  $\left(\bl3l\sigma_{\mu\nu}\lr[\tau]\right)\left(\blr[b]\sigma^{\mu\nu}\q3l\right)$  &  $b\to c\tau\overline{\nu}$ \\ \hline
  $\mathcal{O}^{q}_{T2}$ &  $\left(\bl3l\sigma_{\mu\nu}\lr[\tau]\right)\left(\blr[b]\sigma^{\mu\nu}\ql[i]\vqd[i]\right)$  & $b\to c\tau\overline{\nu}$, $b\to s\tau\overline{\tau}$ \\ \hline
 $\mathcal{O}^{q}_{T3}$ &  $\left(\blep[i]\vl^i \sigma_{\mu\nu}\lr[\tau]\right)\left(\blr[b]\sigma^{\mu\nu}\q3l\right)$  &   $b\to c\tau\overline{\nu}$  \\ 
\tabucline[1.5pt]{-}
\end{tabu}}
\caption{Semi-leptonic four-fermion operators, with leptonic right-handed and scalar currents, and at most one lepton and/or one quark spurion.}
\label{table:sl_2}
\end{table}
 
\begin{table}[t]
\centering
 {\def\arraystretch{1.5}
\begin{tabu}{c|[1.5pt]c|[1.5pt]l}
&Operator                                                                                                                        &    Relevant low-energy processes                                             \\ \tabucline[1.5pt]{-}
$\mathcal{O}^{\ell}_{01}$ &  $\left(\bl3l\gamma^{\mu}\l3l\right)\left(\bl3l\gamma_{\mu}\l3l\right)$  &  --- (flav. cons. leptonic curr.) \\ \hline
$\mathcal{O}^{\ell}_{02}$& $\left(\bl3l\sigma^{a}\gamma^{\mu}\l3l\right)\left(\bl3l\sigma_{a}\gamma_{\mu}\l3l\right)$ &  --- (flav. cons. leptonic curr.) \\ \hline
$\mathcal{O}^{\ell}_{03}$ &  $\left(\bl3l\gamma^{\mu}\l3l\right)\left(\blep[i]\gamma_{\mu}\lep[i]\right)$  &  --- (flav. cons. leptonic curr.) \\ \hline
$\mathcal{O}^{\ell}_{04}$& $\left(\bl3l\sigma^{a}\gamma^{\mu}\l3l\right)\left(\blep[i]\sigma_{a}\gamma_{\mu}\lep[i]\right)$ & $\tau\to\ell\nu\overline{\nu}$  \\ 
\tabucline[1.5pt]{-}
$\mathcal{O}^{\ell}_{11}$ & $\left(\bl3l\gamma^{\mu}\l3l\right)\left(\bl3l\gamma_{\mu}\lep[i]\vld[i]\right)$ & $\tau\to\ell\nu\overline{\nu}$ \\ \hline
$\mathcal{O}^{\ell}_{12}$ & $\left(\bl3l\sigma^{a}\gamma^{\mu}\l3l\right)\left(\bl3l\sigma_{a}\gamma_{\mu}\lep[i]\vld[i]\right)$ & $\tau\to\ell\nu\overline{\nu}$ \\ \hline
$\mathcal{O}^{\ell}_{13}$ & $\left(\bl3l\gamma^{\mu}\lep[i]\vld[i]\right)\left(\blep[j]\gamma_{\mu}\lep[j]\right)$ & $\tau\to\ell\nu\overline{\nu}$, $\tau\to\ell'\ell\overline{\ell}$ \\ \hline
$\mathcal{O}^{\ell}_{14}$ & $\left(\bl3l\sigma^{a}\gamma^{\mu}\lep[i]\vld[i]\right)\left(\blep[j]\sigma_{a}\gamma_{\mu}\lep[j]\right)$ & $\tau\to\ell\nu\overline{\nu}$, $\tau\to\ell'\ell\overline{\ell}$  \\ 
\tabucline[1.5pt]{-}
$\mathcal{O}^{\ell}_{R1}$ & $\left(\bl3l\gamma^{\mu}\l3l\right)\left(\overline{E}_j\gamma_{\mu}{E^j}\right)$ &  --- (flav. cons. leptonic curr.)  \\ \hline
$\mathcal{O}^{\ell}_{R2}$ & $\left(\bl3l\gamma^{\mu}\lep[i]\vld[i]\right)\left(\overline{E}_j\gamma_{\mu}{E^j}\right)$ &  $\tau\to\ell'\ell\overline{\ell}$ \\  
\tabucline[1.5pt]{-}
$ \mathcal{O}^{\ell}_{S1} $ & $\left(\bl3l{E^j}\right)\left(\overline{E}_j\l3l\right)$ &  --- (flav. cons. leptonic curr.) \\ \hline
$\mathcal{O}^{\ell}_{S2} $ & $\left(\bl3l{E^j}\right)\left(\overline{E}_j\lep[i]\vld[i]\right)$ &  $\tau\to\ell'\ell\overline{\ell}$\\ 
\tabucline[1.5pt]{-}
$\mathcal{O}^{\ell}_{T1}$ & $\left(\bl3l\sigma_{\mu\nu}{E^j}\right)\left(\overline{E}_j\sigma^{\mu\nu}\l3l\right)$ &  --- (flav. cons. leptonic curr.) \\ \hline
$\mathcal{O}^{\ell}_{T2}$ & $\left(\bl3l\sigma_{\mu\nu}{E^j}\right)\left(\overline{E}_j \sigma^{\mu\nu}\lep[i]\vld[i]\right)$ &  $\tau\to\ell'\ell\overline{\ell}$\\ 
\tabucline[1.5pt]{-}
\end{tabu}}
\caption{Four-lepton operators}
\label{tbl:leptonic operators}
\end{table}

\section{Observables}
In this Section, we analyse the main experimental constraints on the operators with at most  one lepton 
spurion listed in the previous Section. These includes the non-vanishing constraints from $R_D$ and  $R_{D^*}$,
and a long series of bounds from $\Delta F=1$ and $\Delta F=2$ processes, and $\tau$ decays.
The discussion of selected observables receiving leading contributions from operators with 
two lepton spurions is postponed to Section~\ref{sect:HO}. Unless otherwise specified, the bounds should
be interpreted as bounds on the $C_i$ at the scale $\Lambda$ (i.e.~neglecting RG corrections between $\Lambda $
and the electroweak scale).

\label{sect:obs}
\subsection{Semi-leptonic $b\rightarrow c$ transitions} 

\subsubsection{$B\rightarrow D\ell\bar{\nu}_{\ell}$} 
From the operators in \Table{table:sl_1}, the effective charged-current Lagrangian describing $b\to c$ semi-leptonic decays with light leptons is:
\begin{equation}
\label{c.c._light_leptons}
\mathcal{L}(b\rightarrow c\ell\overline{\nu}_{\ell})=-\frac{4G_{F}}{\sqrt{2}}V_{cb}\left(1+2\cq{04}+2V_{Q_s}\cq{14}\frac{\ckm{cs}}{\ckm{cb}}\right)(\overline{c}_{L}\gamma^{\mu}b_{L})(\overline{\ell}_{L}\gamma_{\mu}\nu_{\ell L}) .
\end{equation}
Since the structure of the Lagrangian in \eq{c.c._light_leptons} is SM-like, the decay width of the process $B\to D\ell\bar{\nu}_{\ell}$
can simply written as
\begin{equation}
\label{width_BtoDlnu}
\Gamma(B\rightarrow D\ell\nu_{\ell} )=\Gamma^{\rm SM}(B\rightarrow D\ell\nu_{\ell} )^{\rm SM} \left| 1+ \delta_{D} \right|^2~,
\qquad \delta_{D}=2\cq{04}+2V_{Q_s}\cq{14}\frac{\ckm{cs}}{\ckm{cb}}~.
\ee 
Using the SM prediction~$\mathcal{B}(B\to D\mu\overline{\nu}_{\mu})^{\mathrm{SM}}=(2.28\pm0.19)\ 10^{-2}$~\cite{EOS}, 
and the experimental result in Ref.~\cite{Olive:2016xmw}, we derive the bound
\begin{equation}
\label{bound_cc_light_leptons}
\Re\left( \cq{04}+V_{Q_s}\cq{14}\frac{\ckm{cs}}{\ckm{cb}}\right) =-0.008\pm0.025 \, ,
\end{equation}
which is compatible with the hypothesis of negligible NP effects in the light lepton channels.

\subsubsection{$B\rightarrow D^{(*)}\tau\bar{\nu}_{\tau}$}
The effective  Lagrangian relevant to semi-leptonic $b\to c$ decays with $\tau$ leptons in the final state is\footnote{~As anticipated, 
here and in $b\to c\tau\bar\nu_\tau$ we ignore the effects of tensor operators, which cannot be distinguished from those of 
left-handed and scalar operators using the limited set of observables presently available.}
\begin{equation}
\begin{aligned}
\label{charged_tau}
\mathcal{L}(b\rightarrow c\tau\overline{\nu}_{\tau})=-\frac{4G_{F}}{\sqrt{2}}V_{cb}\bigg[&\left(1+2\cq{02}+2V_{Q_s}\cq{12}\frac{\ckm{cs}}{\ckm{cb}}\right)(\overline{c}_{L}\gamma^{\mu}b_{L})(\overline{\tau}_{L}\gamma_{\mu}\nu_{\tau}) \\
&+\left(\cqs1+V_{Q_{s}}\cqs2\frac{\ckm{cs}}{\ckm{cb}}\right)(\overline{c}_{L}b_{R})(\overline{\tau}_{R}\nu_{\tau L})\bigg]~.
\end{aligned}
\end{equation}
Contrary to the light lepton case,  in the $\tau$ channel also the scalar operators $\mathcal{O}^{q}_{S1(2)}$ do appear
and the decay amplitudes (and corresponding differential decay widths) 
cannot be expressed as a simple re-scaling of the SM ones.

Expanding to first order in the NP contributions, the 
$B\rightarrow D^{(*)}\tau\bar{\nu}_{\tau}$  differential decay widths  can be decomposed as
\begin{equation}
\label{width_BtoDtaunu}
\frac{\text{d}\Gamma}{\text{d}q^2}(B\rightarrow D^{(*)}\tau\bar{\nu}_{\tau})=(1+2\Delta)\frac{\text{d}\Gamma}{\text{d}q^2}(B\rightarrow  D^{(*)}\tau\bar{\nu}_{\tau} )_{\rm SM}+\Delta_{\rm S}\frac{\text{d}\Gamma}{\text{d}q^2}(B\rightarrow  D^{(*)}\tau\bar{\nu}_{\tau} )_{\rm VS} +O(C_i^2)~,
\end{equation}
with
\be
\Delta=2~\Re\left( \cq{02}+2V_{Q_s}\cq{12}\frac{\ckm{cs}}{\ckm{cb}}\right)~, \qquad 
\Delta_{\rm S}=\Re\left(\cqs1+V_{Q_s}\cqs2\frac{\ckm{cs}}{\ckm{cb}}\right)~.
\ee
Following Ref.~\cite{Becirevic:2016hea},  the two SM differential decay distribution
can be written as\footnote{~See appendix~\ref{app:A} for the definition of the form factors}
\begin{align}
\frac{\text{d}\Gamma}{\text{d}q^2}(B\rightarrow  D\tau\bar{\nu}_{\tau} )_{\rm SM}=&\frac{G_{F}^2 \sqrt{\lambda } \vert V_{cb}\vert^2
   \left(m_{\tau}^2-q^2\right)^2}{384 \pi ^3 m_{B}^3 q^6} \left[3 f_{0}^2(q^2) m_{\tau}^2
   \left(m_{B}^2-m_{D}^2\right)^2+f_{+}^2 \lambda  \left(m_{\tau}^2+2
   q^2\right)\right]~,\\
\frac{\text{d}\Gamma}{\text{d}q^2}(B\rightarrow  D^{*}\tau\bar{\nu}_{\tau} )_{\rm SM}=& \frac{G_{F}^2 \sqrt{\lambda } \vert V_{cb}\vert^2
   \left(m_{\tau}^2-q^2\right)^2 }{384 \pi ^3m_{B} q^6}\bigg[F_{0}^2 \  m_{B}^2 \left(m_{\tau}^2+2
   q^2\right)\notag\\+&q^2 \left(F_{\perp}^2+F_{\parallel}^2\right) \left(m_{\tau}^2+2 q^2\right)+3
   F_{t}^2 \ m_{B}^2 \ m_{\tau}^2\bigg]~.
\end{align}
The non-standard term $\frac{\text{d}\Gamma}{\text{d}q^2}(B\rightarrow  D^{(*)}\tau\bar{\nu}_{\tau} )_{\rm VS}$ arise from the interference 
between the  left-handed and the scalar operators. Its explicit expression in the $D$ and $D^{*}$ case is
\begin{align}
\frac{\text{d}\Gamma}{\text{d}q^2}(B\rightarrow  D\tau\bar{\nu}_{\tau} )_{\rm VS}=& \ \frac{f_{0}^2 G_{F}^2 \sqrt{\lambda} \ m_{\tau} \vert V_{cb}\vert^2 \left(m_{B}^2-m_{D}^2\right)^2 \left(m_{\tau}^2-q^2\right)^2}{64
   \pi^3 m_{B}^3 q^4 (m_{b}-m_{c})}~, \\
\frac{\text{d}\Gamma}{\text{d}q^2}(B\rightarrow  D^{*}\tau\bar{\nu}_{\tau} )_{\rm VS}=& \ \frac{F_{t}^2 G_{F}^2 \sqrt{\lambda } \ m_{B} \ 
  m_{\tau} \vert V_{cb}\vert^2 \left(m_{\tau}^2-q^2\right)^2}{64 \pi ^3 q^4
   (m_{b}+m_{c})}~.
\end{align}

\begin{figure}[t]
\centering
\includegraphics[angle=0,width=10cm]{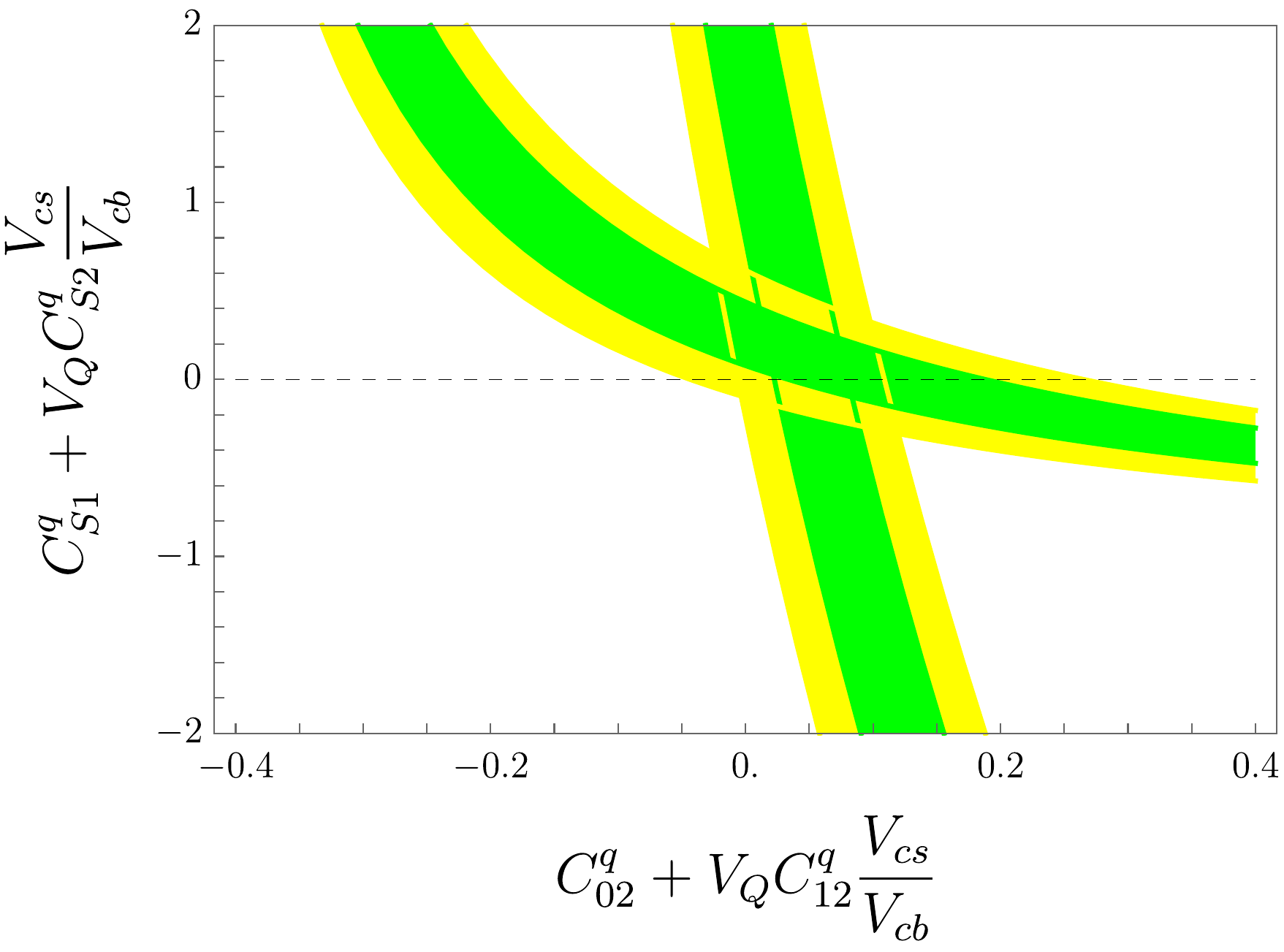}
\caption{Parameter space allowed by the constraint on $\cB(B\to D\tau\overline{\nu}_{\tau})$ and $\cB(B\to D^{*}\tau\overline{\nu}_{\tau})$.
The bands denotes 1 and $2\sigma$ limits (the $C_i$ are assumed to be real).} 
\label{fig:comb_D_Ds}
\end{figure}

In principle, the best discrimination between scalar and left-handed contributions could be obtained by differential measurements of the two spectra, 
using the above formulae. So far these measurements are not available; however, a useful information can be derived also 
comparing the partial widths of two modes. The parameter space allowed by the experimental constraints~\cite{Amhis:2016xyh}  on
 $\cB(B\to D\tau\nu_{\tau})$ and $\cB(B\to D^{*}\tau\nu_{\tau})$   is shown in \fig{fig:comb_D_Ds}. 
 As can be seen, the constraint on the scalar terms is quite weak. Still, it is interesting to note that present data are perfectly 
 compatible with the  absence of scalar terms, while pointing toward a non-negligible modification of the coefficient of the 
 left-handed operator.

As anticipated in the Introduction, the ratios  $R_{D^{(*)}}^{\tau/\mu}$, defined in \eqs{eq:RDexp}{eq:RDSexp}, play a crucial role in our analysis. 
Neglecting scalar terms, as suggested by \fig{fig:comb_D_Ds}, the parameter space allowed by these two ratios 
can easily be derived from \eqs{width_BtoDlnu}{width_BtoDtaunu} and is shown in \fig{rd_rds}. 
If we further  take in account the bound in \eq{bound_cc_light_leptons}, we deduce the following simple relation
\begin{equation}
\Re\left( \cq{02}+V_{Q_2}\cq{12}\frac{\ckm{cs}}{\ckm{cb}} \right)=\frac{1}{4}\left[R_{D^{(*)}}^{\tau/\mu}-1\right],
\end{equation}
which leads us to the following limits
\be
\Re\left(\cq{02}+V_{Q_2}\cq{12}\frac{\ckm{cs}}{\ckm{cb}}\right) ~=~ 
\left\{
\begin{array}{l}
0.06\pm 0.02 \left[R_{D^{*}}\right]~,  \\
0.09\pm 0.04 \left[ R_{D}  \right]~. 
\end{array}
\right.
\label{eq:bounds_rd}
\ee
    
\begin{figure}[t]
\centering
\includegraphics[angle=0,width=8cm]{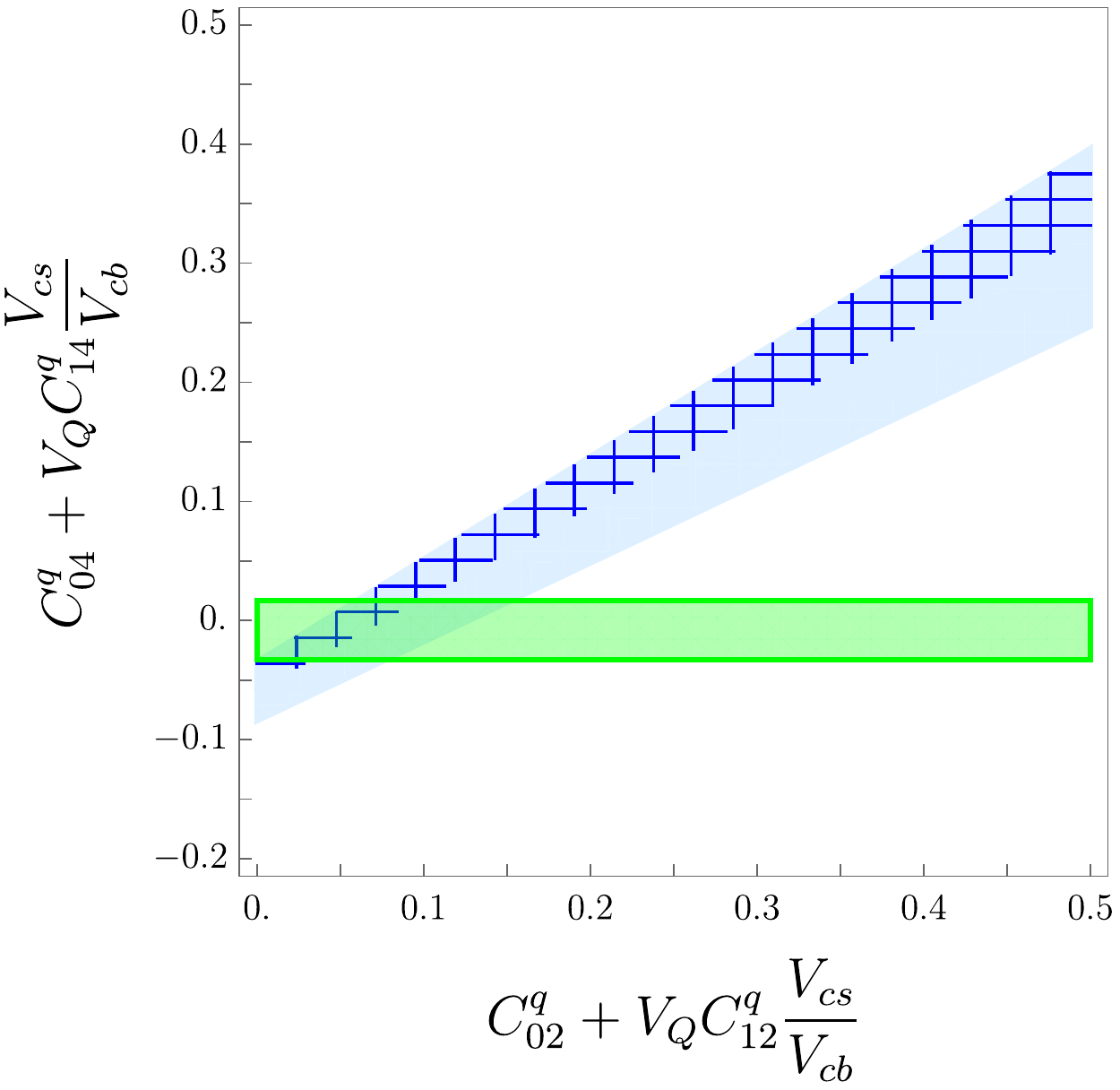}
\caption{Parameter space allowed by the combination of the constraints given by the ratios $R_{D}$ (blue region) and $R_{D^{*}}$ (dashed region) in the hypothesis of negligible scalar current contributions.
 The horizontal band denotes the constraint from $\cB(B\rightarrow D\mu\bar{\nu}_{\ell})$.
}
\label{rd_rds}
\end{figure}
    
\subsection{Semi-leptonic $s\rightarrow u$ transitions} 
The semi-leptonic operators listed in \Table{table:sl_1} generate also  contributions to $s\to u$ transitions
with $\tau$ and light leptons. The relevant effective Lagrangians, taking into account also the SM contributions, 
are
\begin{align}
\mathcal{L}(s\to u\tau\nu) &=-4\frac{G_{F}}{\sqrt{2}}\ckm{us}\left(1+2\cq{06}+2\cq{12} V_{Q_s}\frac{\ckm{ub}}{\ckm{us}}\right)(\overline{u}_{L}\gamma_{\mu}s_{L})(\overline{\tau}_{L}\gamma^{\mu}\nu_{\tau L})\, , \label{Lagrangian:tau to K nu} \\
\mathcal{L}(s\to u\mu\nu) &=-4\frac{G_{F}}{\sqrt{2}}\ckm{us}\left(1+2\cq{08}+2\cq{14}V_{Q_s}\frac{\ckm{ub}}{\ckm{us}}\right)(\overline{u}_{L}\gamma_{\mu}s_{L})(\overline{\mu}_{L}\gamma^{\mu}\nu_{\mu L}) \, . \label{Lagrangian:K to mu nu}
\end{align}
A particularly interesting observable to constrain the NP terms in these Lagrangians is the ratio 
$\mathcal{B}(\tau\to K\nu_{\tau})/\mathcal{B}(K\to\mu\overline{\nu}_{\mu})$, 
where the theoretical uncertainties  on CKM elements and Kaon decay constant cancel out. 
Using the experimental results in \cite{Olive:2016xmw} and the SM input in \cite{Pich:2013lsa} we find
\begin{equation}
R_{sd}^{\tau/\mu}=\frac{\mathcal{B}(K\to\mu\overline{\nu}_{\mu})^{\rm exp}/\mathcal{B}(K\to\mu\overline{\nu}_{\mu})^{\rm SM}}{\mathcal{B}(\tau\to K\nu_{\tau})^{\rm exp}/\mathcal{B}(\tau\to K\nu_{\tau})^{\rm SM}}=1.029\pm0.015~,
\label{eq:Vus_tau}
\end{equation}
which allows us to obtain the following bound
\begin{equation}
\Re\left[\cq{08}-\cq{06}+\left(\cq{14}-\cq{12}\right)V_{Q_s}\frac{\ckm{ub}}{\ckm{us}} \right]=  0.007\pm 0.004~.
\end{equation}
It is worth to stress that $R_{sd}^{\tau/\mu}$ or, equivalently, the comparison of the  $|V_{us}|$  determination  
 from $\tau$ vs.~$K$ decays is nothing but a test of LFU. Interestingly enough, present 
 data exhibits a small tension with the SM prediction also in this case.

\subsection{$\Delta F=2$ processes}
According to the operators in \Table{tbl:quark mixing operators}, the effective Lagrangian relevant to $\Delta F=2$ 
processes is 
\begin{equation}
\mathcal{L}^{\mathrm{NP}}_{\Delta F=2}=-\frac{4G_{F}}{\sqrt{2}}\left(\cqq{01}+\cqq{02}\right)  (V^*_{Q_i})^2 \left[\left(\overline{b}_{L}\gamma^{\mu}d^{i}_{L}\right)^2+\left(\ckmd{ki}\ckm{j3}\right)^2\left(\overline{u}_{L}^{j}\gamma^{\mu}u_{L}^{k}\right)^2\right] .
\label{eq:bbbar_mixing}
\end{equation}
Since the structure of the effective operators is the same as in the SM, 
we can conveniently encode all the NP effects via the ratios
\begin{equation}
R_{B_{q}}^{\Delta F=2}=\frac{\mathcal{A}\left(B_{q}\to\overline{B}_{q}\right)_{\mathrm{SM+NP}}}{\mathcal{A}\left(B_{q}\to\overline{B}_{q}\right)_{\mathrm{SM}}}  \qquad {\rm and }   \qquad
R_{D}^{\Delta F=2}=\frac{\mathcal{A}(D^0\to\overline{D}^0)_{\mathrm{SM+NP}}}{\mathcal{A}(D^0\to\overline{D}^0)_{\mathrm{SM}}}~.
\end{equation}
In the $B$-physics case we find 
\begin{equation}
R_{B_{q}}^{\Delta F=2} =1+\frac{\left(\cqq{01}+\cqq{02}\right) }{  R_{\mathrm{SM}}^{\mathrm{loop} }} \left( \frac{
 V^*_{Q_q} }{ \ckmd{tb}\ckm{tq}} \right)^2~,
\end{equation}
where\footnote{~For analytic and numerical values of $S_{0}(x_{t})$ ans $\eta_{B}$ we refer to Ref.~\cite{Buchalla:1995vs}.}
\begin{equation}
R_{\mathrm{SM}}^{\mathrm{loop}}=\frac{\alpha_{em}}{16\pi s_{w}^2}S_{0}(x_{t})\eta_{B}\approx 1.6\times 10^{-3}~.
\end{equation}
Given the flavor structure of $V_{Q_i}$, we get  very similar bounds from $B_d$ and $B_s$ mixing,
while the bound from $D^0$ is weaker. In particular,  
from the constraint $R_{B_{s}}\in\left(0.86,1.26\right)$~\cite{Barbieri:2014tja} we derive the bound
\begin{equation}
 | V_{Q_1} |^2 | \cqq{01}+\cqq{02} |  <  6.7 \times  10^{-7}~.
\end{equation}
\subsection{FCNC  $b\to s$  transitions}
\label{sect:FCNCs}
\subsubsection{$B\to K^{(*)} \mu\bar\mu$ }
The Lagrangian that encodes FCNC $b\to s$ transition for the light lepton channels is
\begin{equation}
\label{eq:neutral_currents}
\mathcal{L}(b\rightarrow s\ell\overline{\ell}) =-\frac{2G_{F}}{\sqrt{2}}\frac{\alpha_{e}}{2\pi}V_{ts}^* V_{tb}\left[\left(C_{9}+\Delta C_{9}\right)\mathcal{O}_{9}+\left(C_{10}+\Delta C_{10}\right)\mathcal{O}_{10}\right]~,
\end{equation}
where $\mathcal{O}_{9}$ and $\mathcal{O}_{10}$ are defined in \eq{FCNC_wilson coefficients} of 
Appendix~\ref{app:B}, and the shifts of $C_{9}$ and $C_{10}$ in term of the NP Wilson coefficients have the following form:
\begin{align}
\Delta C_{9}=&\frac{2\pi (\cq{13}+\cq{14}+\cqr5)V_{Q_s}}{\alpha_{e}V_{ts}^{*}V_{tb}}~,  &   \Delta C_{10}=&-\frac{2\pi(\cq{13}+\cq{14}-\cqr5)V_{Q_s}}{\alpha_{e}V_{ts}^{*}V_{tb}}~.
\end{align}
Inverting the relations above, we obtain an expression for the two combination of Wilson coefficients that appear in these channels 
as a function of the shifts $\Delta C_{9}$ and $\Delta C_{10}$. These shifts have been constrained 
in Ref.~\cite{Descotes-Genon:2015uva,Altmannshofer:2015sma,Hurth:2016fbr} 
from global fits of various $b\to s \mu\overline{\mu}$ observables (dominated by $B\to K^{*} \mu\bar\mu$ and $B\to K  \mu\bar\mu$ data).
Considering in particular the results
in \cite{Descotes-Genon:2015uva}, namely  $\Delta C_{9}=-1.05\pm 0.35$ and $\Delta C_{10}=0.3\pm 0.4$, we find
\begin{align}
\Re\left[ (\cq{13}+\cq{14}+\cqr5)V_{Q_s} \right]&=(-4.9\pm 1.7) \times 10^{-5}~, \\
\Re\left[ (\cq{13}+\cq{14}-\cqr5)V_{Q_s} \right]&=(1.4\pm 1.9) \times 10^{-5}~.
\end{align}

\subsubsection{$B\rightarrow K^{(*)}\tau\overline{\tau}$}
In principle, $b\to s \tau\overline{\tau}$ transitions would be excellent probes of our EFT construction.  
However, the current experimental bounds~\cite{TheBaBar:2016xwe} are too weak to draw significant constraints. 
For completeness, and in view of future data, we report here the relevant formulae. 

The relevant effective Lagrangian   can be expressed as 
\begin{eqnarray}
\label{btok_massive_leptons_lagr}
\mathcal{L}(b\rightarrow s\tau\overline{\tau}) =-\frac{2G_{F}}{\sqrt{2}}\frac{\alpha_{e}}{2\pi}V_{ts}^* V_{tb}\bigg[\left(C_{9}+\Delta C_{9}^\tau\right)\mathcal{O}_{9}  +\left(C_{10}+\Delta C_{10}^\tau\right)\mathcal{O}_{10} 
+C_{S}^\tau(\mathcal{O}_{S}-\mathcal{O}_{P})\bigg],
\end{eqnarray}
where the operators $\mathcal{O}_{9}$, $\mathcal{O}_{10}$, $\mathcal{O}_{S}$ and $\mathcal{O}_{P}$ 
are defined in \eq{FCNC_wilson coefficients} with the identification $\ell=\tau$.
In terms of the Wilson coefficients of the operators in \Tables{table:sl_1}{table:sl_2}, the NP contributions are given by
\begin{align}
\Delta C_{9(10)}^\tau= & \pm \frac{2\pi(\cq{11}+\cq{12}\pm \cqr2)V_{Q_s}}{\alpha_{e}V_{ts}^{*}V_{tb}}~,   & C_{S}^\tau=&\frac{2\pi \cqs2 V_{Q_s}}{\alpha_{e}V_{ts}^{*}V_{tb}}~.
\end{align}

\subsubsection{$B\to K^{(*)} \nu\overline{\nu}$}
From the operators in \Table{table:sl_1} we get the following Lagrangian for $b\to s\nu\overline{\nu}$ transitions 
\begin{equation}
\label{btosnunu_lagr}
\mathcal{L}(b\to s\nu\overline{\nu})  =-\frac{2G_{F}}{\sqrt{2}}\frac{\alpha_{e}}{2\pi}\left[\sum_{\ell=e,\mu}(C_{\nu}+\Delta C_{\nu_\ell})\mathcal{O}_{\nu_{\ell}}+(C_{\nu}+\Delta C_{\nu_{\tau}})\mathcal{O}_{\nu_{\tau}}\right]
\end{equation}
where the operators $\mathcal{O}_{\nu_{\ell}}$ and $\mathcal{O}_{\nu_{\tau}}$ are defined starting from those in 
\eq{FCNC_wilson coefficients}  as 
\be
\mathcal{O}_{\nu_{\ell} (\nu_{\tau})}= \left. \mathcal{O}_{9}-\mathcal{O}_{10}\right|_{\ell=\nu_{\ell} (\nu_{\tau})}~.
\ee
The shifts of the Wilson coefficients due to NP effects are
\begin{align}
\Delta C_{\nu_{\ell}}&=\frac{2\pi V_{Q_2}(\cq{13}-\cq{14})}{\alpha_{e}V_{ts}^{*}V_{tb}}~,  &   \Delta C_{\nu_{\tau}}&=\frac{2\pi V_{Q_2}(\cq{11}-\cq{12})}{\alpha_{e}V_{ts}^{*}V_{tb}}~.
\end{align}
Since the Lagrangian in \eq{btosnunu_lagr} has a SM-like structure, the differential decay widths for 
$B\to K^{(*)} \nu\overline{\nu}$ decays can be expressed as 
\begin{equation}
\frac{\mathrm{d}\Gamma}{\mathrm{d}q^2}(B\to K^{(*)} \nu\overline{\nu})=\frac{\mathrm{d}\Gamma}{\mathrm{d}q^2}(B\to K^{(*)} \nu\overline{\nu})^{\rm SM} \left| 1+\frac{2}{3}\frac{\Delta C_{\nu_{\ell}}}{C_{\nu}}+\frac{1}{3}\frac{\Delta C_{\nu_{\tau}}}{C_{\nu}}\right|^2~.
\end{equation}
In the case case, the SM spectrum can be be read from  \eq{width_BtoKtautau} 
setting $C_{9}=-C_{10}=C_{\nu}$, $C_{S}=$0 and $m_{\ell}=0$.
Using the SM $C_{\nu}= -6.35$~\cite{Brod:2010hi} and the hadronic form factors 
in \cite{Bouchard:2013pna}, from the experimental bound in \cite{Olive:2016xmw} we obtain
\begin{equation}
-9.3 <  \Re\left( \frac{\Delta C_{\nu_{\tau}}}{C_{\nu}} \right) <  3.3 \, , 
\end{equation}
in the limit  $|\Delta C_{\nu_{\tau}}| \ll  |\Delta C_{\nu}|$. This implies in turn
\begin{equation}
\Re\left[  V_{Q_s}(\cq{11}-\cq{12}) \right] = (0.9 \pm  1.8 ) \times 10^{-3}~. 
\end{equation}

\subsection{Leptonic $\tau$ decays}

\subsubsection{$\tau\to\ell\nu\overline{\nu}$}
\label{sect:taulnu}
The effective Lagrangian generating  $\tau\to\mu\nu\bar{\nu}$ decay amplitudes  at the tree level is 
\begin{equation}
\begin{aligned}
\label{tau_to_lepnunu_lagr}
\mathcal{L}(\tau\to\mu\nu\bar{\nu})=-\frac{4G_{F}}{\sqrt{2}}\lbrace&(1+2\cl04)(\overline{\nu}_{\tau_L}\gamma^{\lambda}\tau_{L})(\overline{\mu}_{L}\gamma_{\lambda}\nu_{\mu_L})+V_{L_{\mu}}[(3\cl12-\cl11)(\overline{\nu}_{\tau_L}\gamma^{\lambda}\tau_{L})(\overline{\mu}_{L}\gamma_{\lambda}\nu_{\tau_L})  \\
+&(3\cl14-\cl13)(\overline{\nu}_{\mu_L}\gamma^{\lambda}\tau_{L})(\overline{\mu}_{L}\gamma_{\lambda}\nu_{\mu_L}) +(\cl14-\cl13)(\overline{\nu}_{e_L}\gamma^{\lambda}\tau_{L})(\overline{\mu}_{L}\gamma_{\lambda}\nu_{e_L})]\rbrace .
\end{aligned}
\end{equation}
Since the interacting structure is the same occurring within the SM,  the decay width can be simply written as 
\begin{equation}
\Gamma(\tau\to\mu\nu\overline{\nu})= \Gamma^{\rm SM}(\tau\to\mu\nu_{\tau}\bar{\nu}_{\mu}) \times 
\left| 1+2\cl04 \right|^2~,
\label{eq:Gtaufact}
\end{equation}
where $\Gamma^{\rm SM}(\tau\to\mu\nu_{\tau}\bar{\nu}_{\mu})$ is given in \cite{Pich:2013lsa}.
We can now consider the observable $R_{\tau}^{\tau/\ell}$, defined as
\begin{equation}
R_{\tau}^{\tau/\ell_{1,2}}=\frac{\mathcal{B}(\tau\to\ell_{2,1}\nu\bar\nu)_{\mathrm {exp}}/\mathcal{B}(\tau\to\ell_{2,1}\nu\bar\nu)_{\mathrm {SM}}}{\mathcal{B}(\mu\to e\nu\bar\nu)_{\mathrm {exp}}/\mathcal{B}(\mu\to e\nu\bar\nu)_{\mathrm {SM}}}~,
\end{equation}
whose value can be extracted from \cite{Amhis:2016xyh}:
\begin{align}
R_{\tau}^{\tau/\mu}&=1.0020\pm 0.0030~,   &  R_{\tau}^{\tau/e}&=1.0058\pm 0.0030 \, .
\label{eq:Rtaumu}
\end{align}
This allows us to constrain with very good precision $\Re(\cl04)$. 

Given the strength of these constraints (that affect a combination of $C_i$ not parametrically suppressed by spurions),
in this case it is necessary to take into account also the effect of radiative corrections~\cite{Feruglio:2016gvd}.
The latter are identical for SM and NP amplitudes below the electroweak scale, i.e.~they factorise in Eq.~(\ref{eq:Gtaufact}).
This implies that we can directly translate the experimental 
bounds~(\ref{eq:Rtaumu}) into a constraint on  $\Re(\cl04)$ renormalised at the electroweak scale:
\begin{equation}
\Re\left[ \cl04\left(M_W\right) \right]    = (5\pm 7)\times 10^{-4} ~.
\label{eq:tauconstr}
\end{equation}
On the contrary, radiative corrections are different for SM and NP amplitudes  above the electroweak scale.
In particular, a sizeable contribution to $\tau\to\mu\nu_{\tau}\bar{\nu}_{\mu}$ is generated by
the semi-leptonic operators  contributing to $R_{D^{(*)}}$. To a first approximation, this effect 
can be taken into account by the leading contribution to the RG evolution of  $\cl04$~\cite{Feruglio:2016gvd}
\begin{equation}
\cl04\left(M_{W}\right)=\cl04\left(\Lambda\right)+\frac{3y_{t}^2}{8\pi^2}\vert V_{tb}\vert^2\left[ 
\cq{02}\left(\Lambda\right)+V_{Q}\cq{12}\left(\Lambda\right) \right]\times \left[\log\left(\frac{\Lambda^2}{m_{t}^2}\right)+\frac{1}{2}\right]\, .
\end{equation}
Using this result, and setting $\Lambda\approx 1~\mathrm{TeV}$, the constrain in \eq{eq:bounds_rd} becomes
\begin{equation}
 \Re\left[\cl04(\Lambda) \right]=- \left( 1.2\pm 0.09_{\tau} \pm 0.53_{R_D}\right)\times 10^{-2} \ ,
 \end{equation}
where we have explicitly separated the small error due to Eq.~(\ref{eq:tauconstr}) and the sizable error due to the 
input value of $\cq{02}(\Lambda)$ or, equivalently, due to $R_{D^{(*)}}$. 
The fact that we need a non-vanishing value for $\cl04(\Lambda)$ in order to  cancel the large NP contribution generated by 
$\cq{02}(\Lambda)$ necessarily signals  a fine tuning in the EFT. The minimum amount of this fine-tuning is $\approx 10\%$,
that is what we deduce comparing the central value of $\cl04(\Lambda)$ with the error determined by  Eq.~(\ref{eq:tauconstr}).
The fine-tuning would increase if the central value of $\cl04(\Lambda)$ were not natural. 
However, this 
can be avoided with the power-counting scheme that we will introduce in Section~\ref{sect:EFTexp}.

\subsubsection{$\tau\rightarrow\ell\overline{\ell}\ell'$}
The purely leptonic  LFV  decays $\tau\rightarrow\ell\overline{\ell}\ell'$, which are highly suppressed in the SM, 
arises naturally in our framework due to the operators $\mathcal{O}^{\ell}_{13}$ and $\mathcal{O}^{\ell}_{14}$ in \Table{tbl:leptonic operators}. The corresponding effective Lagrangian is:
\begin{equation}
\begin{aligned}
\label{tau_to_3lep_lagr}
\mathcal{L}(\tau\to\ell^{i}\ell\overline{\ell})=-\frac{4G_{F}}{\sqrt{2}}\bigg[&(\cl13+\cl14)V_{L}^{i}(\overline{\ell}^{i}_{L}\gamma_{\mu}\tau_{L})(\overline{\ell}_{L}\gamma^{\mu}\ell _{L})+\clr2V_{L}^{i}(\overline{\ell}^{i}_{L}\gamma_{\mu}\tau_{L})(\overline{\ell}_{jR}\gamma^{\mu}\ell _{R}) \\
&+\clt2 V_{L}^{i}(\bar{\ell}_{R}\sigma_{\mu\nu}\tau_{L})(\bar{\ell}^{i}_{L}\sigma^{\mu\nu}\ell _{R})\bigg].
\end{aligned}
\end{equation}
In the $\tau\to\mu e\bar e$ case  we get
\begin{equation}
\Gamma(\tau\to\mu e\overline{e})=\left(\vert \cl13+\cl14\vert^2 +\vert \clr2\vert^2+\vert\clt2\vert^2\right) \vert V_{L}\vert^2\tilde{\Gamma}(\tau\to \mu e\bar e)~,
\end{equation}
where $\tilde{\Gamma}(\tau\to\mu e\bar e)=\Gamma(\tau\to\mu\bar{\nu}\nu)$ in the limit $m_e\to0$. 
From the experimental bound $\mathcal{B}(\tau\to\overline{e}e \mu)^{\rm exp}<1.8\times 10^{-8}$~\cite{Olive:2016xmw} we obtain
\begin{equation}
 \vert V_{L}\vert\sqrt{\left(\vert \cl13+\cl14\vert^2 +\vert \clr2\vert^2+\vert\clt2\vert^2\right)}  < 3.2\times 10^{-4}~.
\end{equation}
An almost identical bound is obtained from  $\mathcal{B}(\tau\to 3 \mu)^{\rm exp}<2.1 \times 10^{-8}$.

\subsection{Semi-leptonic LFV transitions}

\subsubsection{$B \to \tau\overline{\mu}$}
The leading contributions to the semi-leptonic  LFV $b\to d\tau\mu$ transitions can be computed 
in terms of the following effective Lagrangian
\begin{equation}
{\mathcal L}^{\mathrm{NP}}(b \to d\tau\overline{\mu} ) =  - \frac{{4{G_F}}}{{\sqrt 2 }}\left( \cq{31} +\cq{32} \right)V_{Q_d}\vl({\overline d _L}{\gamma ^\mu }{b_L})({\overline \tau  _L}{\gamma _\mu }{\mu _L})~,
\end{equation}
that in the $B\to\tau\mu$ case leads to
\begin{equation}
\Gamma \left( B \to \tau\overline{\mu} \right) = {\left( \cq{31} + \cq{32} \right)^2}\left| \vq \right|^2\left| \vl \right|^2\frac{{G_F^2f_{{B}}^2{\sqrt{\lambda(m_{B}^2+m_{\tau}^2+m_{\mu}^2) }}}}{{8\pi m_{{B}}^3}}\left[ m_{B} ^2\left( {m_\tau ^2 + m_\mu ^2} \right) - {{\left( {m_\tau ^2 - m_\mu ^2} \right)}^2} \right]~.
\end{equation}
Using $f_{B}=(207^{+17}_{-9}) \ \mathrm{MeV}$~\cite{Gelhausen:2013wia} and the current experimental bound 
$\cB(B\to\tau\mu) < 2.2 \times 10^{-5}$~\cite{Olive:2016xmw} we obtain
\begin{equation}
\vert \cq{31}+\cq{32}\vert\vert \vl V_{Q_d}\vert < 1.8\times 10^{-3}~.
\end{equation}

\subsubsection{$\tau \to   \mu \omega  $ and $\tau \to\mu \rho$}
Semi-leptonic  LFV transitions can occur in $\tau$ decays via 
the following effective Lagrangian
\begin{equation}
\mathcal{L}\left(\tau\to\mu V \right)=-\frac{4G_{F}}{\sqrt{2}}\vl\left[(\cq{23}-\cq{24})\left(\overline{u}_{L}\gamma^{\mu}u_{L}\right)+(\cq{23}+\cq{24})\left(\overline{d}_{L}\gamma^{\mu}d_{L}\right)\right]\left(\overline{\tau}_{L}\gamma_{\mu}\mu_{L}\right) \, .
\end{equation}
The two most interesting cases are $V=\rho$ and $V=\omega$, which allow us to constrain separately the Wilson coefficients $\cq{23}$ and $\cq{24}$. The decay widths of these two processes are:
\begin{equation}
\begin{aligned}
\Gamma\left(\tau\to\mu\rho\right)&=\frac{G_{F}^2}{8\pi}\vert\cq{24}\vert^2 \vert\vl\vert^2 f_{\rho}^2\frac{\sqrt{\lambda(m_{\tau}^2, m_{\rho}^2 ,m_{\mu}^2)}}{m_{\tau}^3}\left[(m_{\tau}^2- m_{\mu}^2)^2+m_{\rho}^2(m_{\tau}^2+m_{\mu}^2-2m_{\rho}^2)\right] \, ,\\
\Gamma\left(\tau\to\mu\omega\right)&=\frac{G_{F}^2}{8\pi}\vert\cq{23}\vert^2 \vert\vl\vert^2 f_{\omega}^2\frac{\sqrt{\lambda(m_{\tau}^2, m_{\omega}^2 ,m_{\mu}^2)}}{m_{\tau}^3}\left[(m_{\tau}^2- m_{\mu}^2)^2+m_{\omega}^2(m_{\tau}^2+m_{\mu}^2-2m_{\omega}^2)\right]\, .
\end{aligned}
\end{equation}
Using the decay constant for both $\omega$  and $\rho$ mesons in \cite{Alte:2016yuw} and the experimental bounds in \cite{Olive:2016xmw} 
we get the following limits
\begin{align}
\vert\cq{24}\vert \vert \vl\vert& <1.4 \times 10^{-4}  \qquad {\rm from} \qquad \cB(\tau\to\mu\rho) < 1.8\times 10^{-8} ~,  \\
\vert\cq{23}\vert \vert \vl\vert& <3.2 \times 10^{-4} \qquad {\rm from} \qquad \cB(\tau\to\mu\omega) <  4.7 \times 10^{-8}~. 
\end{align}

\subsubsection{$\Upsilon \to \tau\overline{\mu}$ and $\eta_b \to \tau\overline{\mu}$}
As listed in \Tables{table:sl_1}{table:sl_2}, in principle LFV decays of $b\overline{b}$ bound states are also possible. 
The Lagrangian relevant to these processes is
\begin{equation}
\label{Lagrangian_bbbar}
\mathcal{L}(b\to b\tau\mu)=-\frac{4G_{F}}{\sqrt{2}}\vl\left[(\cq{21}+\cq{22}) \left(\overline{b}_{L}\gamma^{\mu}b_{L}\right)\left(\overline{\tau}_{L}\gamma_{\mu}\mu_{L}\right)+\cqs{3}(\overline{b}_{L}b_{R})(\overline{\tau}_{R}\mu_{L})\right] \, .
\end{equation}
In the $\Upsilon\to\tau\mu$ case we find
\begin{equation}
\Gamma\left(\Upsilon\to\tau\mu\right)=\frac{G_{F}^2}{24\pi}\vert\cq{21}+\cq{22}\vert^2\vert\vl\vert^2 f_{\Upsilon}^2\frac{\sqrt{\lambda(m_{\Upsilon}^2, m_{\tau}^2 ,m_{\mu}^2)}}{m_{\Upsilon}^3}\left[2m_{\Upsilon}^4-m_{\Upsilon}^2(m_{\tau}^2+m_{\mu}^2)-(m_{\tau}^2-m_{\mu}^2)^2\right] \, .
\end{equation}
From the experimental bound $\mathcal{B}(\Upsilon \to \tau\overline{\mu}) < 6 \times 10^{-6}$~\cite{Olive:2016xmw}, 
using $f_{\Upsilon}=(684.4 \pm 4.6) \ \mathrm{MeV}$~\cite{Alte:2016yuw}, we get 
\begin{equation}
\label{bound_upsilon}
\vert \cq{21}+\cq{22}\vert \vert\vl\vert <0.52 \, .
\end{equation}
The  bound in \eq{bound_upsilon} is significantly weaker than all LFV bounds discussed so far, despite the
stringent experimental limit on $\mathcal{B}(\Upsilon \to \tau\overline{\mu})$. This is trivial consequence of the fact,
contrary to  $\tau$ and  $B$ mesons, the $\Upsilon$ does not decay via weak interactions.
It is then easy to verify that the constraints following from the  $O(1\%)$ experimental bound on $\mathcal{B}(\eta_{b}\to\mu\bar\mu)$
are irrelevant.

\section{Consistency of the EFT construction}
\label{sect:Consistency}

\subsection{Power-counting scheme}
\label{sect:EFTexp}

\begin{table}[t]
\begin{center}
\centering
\renewcommand{\arraystretch}{1.4} 
\hglue -0.4 true cm 
\begin{tabu}{c|[1.5pt]c|[1.5pt]c|[1.5pt]c|l}
\multirow{2}{*}{Process} & \multirow{2}{*}{Combination}  & \multirow{2}{*}{Constraint}  &   Parametric     & Order of \\  [-5pt] 
&     &  &  scaling  & magnitude \\ 
 \tabucline[1.5pt]{-}
$R_{D^{(*)}}$ & $\Re\left(\cq{02}+V_{Q_s}\cq{12}\frac{\ckm{cs}}{\ckm{cb}}\right)$ & $0.09\pm 0.04$ & $1$  & $    10^{-1}$  \\	
\tabucline[1.5pt]{-}
$B\to D\mu\nu_{\mu}$ &  $\Re\left(\cq{04}+V_{Q_s }\cq{14}\frac{\ckm{cs}}{\ckm{cb}}\right)$ & $- (0.8\pm 2.5) \times 10^{-2} $ & $\left(\epsl{L}\right)^2$
& $     10^{-2}$
 \\ \hline
$\tau\to\mu\nu\overline{\nu}$ & $\Re\left(\cl04\right)$ & 
$  - (1.2 \pm 0.5 )\times 10^{-2} $   & $  \left(\epsl{L}\right)^2  r_{q\ell} $ & 
$    10^{-2}~ r_{q\ell}  $
\\  \hline
$R_{sd}^{\tau/\mu}$ &  $\begin{array}{c} \Re\big[\cq{08}-\cq{06}+ \\
(\cq{14}-\cq{12})| V_{Q_s } \ckm{ub}/\ckm{us}| \big] \end{array}  $  &  
  $ (0.7\pm 0.4)\times 10^{-2} $  
 & $\left(\epsq{L}\right)^2$
& $\leq  10^{-2}$ \\
 \tabucline[1.5pt]{-} 
   $\begin{array}{c} \tau\to\mu e e \\   \tau\to 3\mu  \end{array}$  &  $\begin{array}{c} \vert V_{L} \vert \times \big(\vert \cl13  + \cl14\vert^2  + \\
+ \clr2\vert^2+\vert\clt2\vert^2 \big)^{1/2}  \end{array} $ &  $\leq 3.2\times 10^{-4}$  & 
 $\epslp\left(\epsl{L,R}\right)^2 r_{q\ell} $   &  $10^{-3}  \left(\frac{  \epslp}{0.1} \right)   r_{q\ell} $    \\ \hline 
$\tau\to\rho\mu$ & $\vert\cq{24}\vert \vert \vl\vert$& $\leq 1.4\times 10^{-4}$ & $\epslp (\epsq{L})^2$ & 
 $\leq  10^{-3} \left(\frac{  \epslp}{0.1} \right)$  \\ \hline
$\tau\to\omega\mu$ & $\vert\cq{23}\vert \vert \vl\vert$& $\leq 3.2\times 10^{-4}$ & $\epslp (\epsq{L})^2$  & 
 $\leq  10^{-3} \left(\frac{  \epslp}{0.1}\right)$  \\ 
 \tabucline[1.5pt]{-} 
    $B\to K \nu\bar{\nu} $   &     $\Re(  \cq{11}-\cq{12} ) $   &  $ (2.2 \pm 4.5) \times 10^{-2}$ & $ \epsqp $    &  $  10^{-2} 
   \left( \frac{  \epsqp}{0.1} \right)  $  \\  \hline
 $B^{0}-\overline{B}^{0}$ & $ \left|\cqq{01}+\cqq{02} \right| $ &  $\leq 0.42 \times 10^{-3}    $ & $\left(\epsqp\right)^2  r_{q\ell}^{-1}  $  
 &  $  10^{-3} \left(\frac{  \epsqp}{0.1} \right)^{2}   r_{q\ell}^{-1} $ 
  \\  \hline
\multirow{2}{*}{$B\to K^{(*)}\mu\bar\mu$} 	
& $\Re\left(\cq{13}+\cq{14}\right) $ & $-(0.8\pm 0.3)\times 10^{-3} $ & $\epsqp\left(\epsl{L }\right)^2$ 
& \!\!\!\!  \multirow{2}{*}{ $  10^{-3} \left( \frac{  \epsqp}{0.1} \right) $}    \\
& $\Re\left( \cqr5 \right)$ &  $-(0.4\pm 0.3)\times 10^{-3}$ & $\epsqp\left(\epsl{ R}\right)^2$\\ \hline
$B_{d}\to\tau\mu$ & $\vert\cq{31}+\cq{32} \vert $ & $\leq 4.5\times 10^{-2}$ & $\epsqp\epslp$ & 
 $    10^{-3}   \left( \frac{  \epsqp\epslp }{10^{-2}} \right) $ 
\\ \hline
\tabucline[1.5pt]{-}
\end{tabu}
\renewcommand{\arraystretch}{1}
\caption{Most relevant constraints on the Wilson coefficients, as obtained in Section~\ref{sect:obs}.
In the last two columns we report the parametric scaling  of the (leading) Wilson coefficients,
according to  the rules defined in Section~\ref{sect:EFTexp}, and the  order of magnitude 
following from the  overall EFT scale and the choice of the $\epsilon_i$ 
reported in Eqs.~(\ref{eq:scale})--(\ref{eq:counting}).}
\label{tbl:results}
\end{center}
\end{table}

We are now ready to discuss the consistency of the EFT construction for the 
leading four-fermion operators listed in Section~\ref{sect:basis}.
The constraints on the Wilson coefficients 
obtained by comparison with data, as discussed in Section~\ref{sect:obs}, 
are summarised in Table~\ref{tbl:results}.  
Assuming a non-vanishing value for the combination of $C_i$ contributing to 
$R_{D^{(*)}}$, the construction can be considered  consistent  if we 
are able to justify, via appropriate re-scaling of the fields (motivated by dynamical assumptions), 
the strong suppression of all the other terms in Table~\ref{tbl:results}.  

Inspired by the explicit dynamical models proposed  in the literature,
we assume a generic framework where the NP sector is coupled preferentially 
to third generation SM fermions (i.e.~the ${\mathcal G}_{\rm flavor}$ singlets),
while the  coupling to the light SM fermions are suppressed by 
small mixing angles (as suggested e.g.~in~\cite{Glashow:2014iga,Greljo:2015mma}).
 As a result of this hypothesis, we re-scale the light SM fermion fields as following	 
\be 
Q_{L}^{i}\to\epsq{L}Q_{L}^{i}~, \qquad L^{i}\to\epsl{L} L^{i}~,  \qquad E_{R}^{i}\to\epsl{R}E_{R}~,
\ee
every time these fields appear in bilinear combinations without spurions. Furthermore, given the underlying dynamics
is potentially different in  quark and lepton sectors, we introduce the flavor-blind re-scaling factor 
$r_{q\ell}$, which allow us to enhance (suppress) the relative weight of leptonic (four-quark) 
operators vs.~semi-leptonic ones. 
Finally, as far as the size of the spurions are concerend, we perform the following re-scaling:
\begin{align}
& | \vq | \to \epsqp |\ckm{ts}| & & | \vl |\to\epslp~.
\end{align}
As discussed in Section~\ref{sect:setup}, in absence of a specific alignment of the  $U(2)_q$ singlets
to left-handed bottom or top quarks, we expect $|V_Q |= O(|V_{ts}|)$. 
The parameter $\epsqp$ is thus a measure of the tuning 
in the (quark) flavor space. On the contrary, $\vl$ parametrizes the unknown size of the spurion in the 
lepton sector. 

By construction, the only combination in Table~\ref{tbl:results}  without $\epsilon_i$ suppression  is the one contributing
to $R_{D^{(*)}}$. This allows us to determine the overall scale of the EFT.
From the central value of the $R_{D^{(*)}}$ anomaly we deduce 
\be
\Lambda \approx (0.09)^{-1/2} v^2_F \approx 600~{\rm GeV}
\label{eq:scale}
\ee
or a natural size of $O(10^{-1})$ for the $C_i$ in absence of $\epsilon_i$ factors.  
A non-vanishing NP contribution to $R_{D^{(*)}}$ necessarily implies a non-vanishing value 
for  $\cl04(\Lambda)$ to cancel NP contributions in $\tau\to\mu\nu\overline{\nu}$. As  
discussed in Section~\ref{sect:taulnu}, this fact necessarily implies a fine-tuning of at least $10\%$,
obtained by comparing error and central value of $\cl04(\Lambda)$. 
This fine-tuning does not increase if the central value of $\cl04(\Lambda)$ 
is natural,  that is what we obtain setting $\left(\epsl{L}\right)^2 r_{q\ell} = O(10^{-1})$. 
More generally, we find that all entries in Table~\ref{tbl:results} have the correct order
of magnitude for the following choice of parameters 
\be
\epsl{L}\approx 0.3~,  \qquad \epsq{L}\leq 0.3~, \qquad \epslp \leq 0.1~,  
\label{eq:counting}
\ee
and 
\be
\qquad \epsqp \approx 0.1~,  \qquad r_{q\ell} =O(1)~.
\ee
Using these reference values we determine the numerical scaling reported in the last column of Table~\ref{tbl:results}.
Setting $\epslp=0.1$, that is the preferred value for a natural solution of the $R_K$ anomaly (see Sect.~\ref{sect:HO}),
a residual fine-tuning appears in the operators contributing to LFV $\tau$ decays; however, this tuning 
is less severe that the one occurring in $\cl04$ and the experimental bounds 
can easily be satisfied setting a slightly smaller value for  $\epsq{L}$.

A second significant source of tuning is the one implied by the smallness of $\epsqp$,
that is a necessary consequence of both $\Delta F=2$ and $b\to s$ FCNC constraints.
Given the difference parametric dependence of these constraints from $\epsq{L}$ and $r_{q\ell}$,  is not possible to obtain 
a good fit  to all data for larger values of $\epsqp$. This implies that the EFT requires a non-negligible 
tuning in flavor space, namely a $O(10\%)$ alignment of the $U(2)_q$ singlets 
to left-handed bottom quarks.

We finally address the issue of the stability of this modified power counting scheme
under radiative corrections. Being not associated to spurions of the flavor symmetry,
the value of  $\epsl{L}$ and  $\epsq{L}$ cannot be arbitrarily small. Indeed, even 
if we do not introduce operators with light quarks  at the heavy scale $\Lambda$,
these are radiatively generated at lower scales (as pointed out in Ref.~\cite{Feruglio:2016gvd}). 
On general grounds, for $\Lambda \sim 1$~TeV, we expect the construction 
to be radiatively stable if  
\be
\left( \epsilon^{q(\ell)}_{L} \right)^2 > \frac{N_C}{16\pi^2} \log(\Lambda^2/m_t^2)  \approx 7\%~.
\ee
We have explicitly verified that, adopting the numerical values in Eq.~(\ref{eq:counting}), 
loop contributions compete with initial conditions only in the case of $\cl04$, 
while they are numerically subleading for the other combinations of Wilson coefficients 
in  Table~\ref{tbl:results}.

\subsection{Processes starting at $O(|V_L|^2)$}
\label{sect:HO}

So far we restricted the attention to processes with at most one $V_{L}$ spurion.
A complete analysis of all the operators appearing at $O(|V_{L}|^2)$ is beyond the scope of our analysis.
However, there are two interesting LFU ratios receiving leading contributions at $O(|V_{L}|^2)$ that is worth to analyse
to further tests the consistency of the EFT:  $\RK$  defined in Eq.~(\ref{eq:RKexp}), 
and a similar $\mu/e$ ratio in $\tau \to \ell \nu\bar\nu$ decays. 

\subsubsection{The LFU ratio $\RK$}
The $O(|V_{L}|^2)$ operators generating a breaking of LFU at the tree-level in $b\to s\ell\overline{\ell}$ decays
have the form
\begin{eqnarray}
\mathcal{O}^{q-2}_{13} &=& \left(\bq3l\gamma^{\mu}\ql[i]\vqd[i]\right)\left(V_{Lj} \blep[j] \gamma_{\mu}\lep[i]\vld[i] \right)~,  \\
\qquad
\mathcal{O}^{q-2}_{14} &=& \left(\bq3l\sigma^{a}\gamma^{\mu}\ql[i]\vqd[i]\right)\left(V_{Lj} \blep[j] \sigma_{a}\gamma_{\mu}\lep[i]\vld[i] \right)~.
\end{eqnarray}
Using the notations of Section~\ref{sect:FCNCs}, these would generate the following non-universal 
shift in the $\ell=\mu$ case
\begin{align}
\Delta C_{9}^{\mu}= - \Delta C^{\mu}_{10} = 
\frac{\vert \vq\vert \left[\left(C^{q-2}_{13}\right)_{\mu}+\left(C^{q-2}_{14}\right)_{\mu}\right]}{\frac{\alpha}{2\pi}\vert \ckmd{ts}\ckm{tb}\vert}
= \left(0.8 \times 10^{3}\right) \times O\left[ \epsilon^\prime_q  (\epslp)^2 \right]~,
\end{align}
where on the r.h.s.~we have indicated the parametric scaling as defined in the previous Section.
The central value of $R^{\rm exp}_K$ can be obtained for  $\Delta C_{9}^{\mu}=-\Delta C_{10}^{\mu} 
\approx -1.0$~\cite{Descotes-Genon:2015uva}.
As can be seen, this value can naturally be obtained for  
 $\epsilon^\prime_q \approx \epslp \approx 0.1$, i.e.~in absence of further fine-tuning 
compared to what determined from the leading operators.

\subsubsection{LFU violations in $\tau \to \ell \nu\bar\nu$ decays}

At $O(|V_{L}|^2)$ one can generate a violation of $\mu/e$ universality in $\tau \to \ell \nu\bar\nu$, which is experimentally 
strongly constrained. The relevant operator is
\begin{eqnarray}
\mathcal{O}^{\ell-2}_{04} &=& \left(\bl3l\sigma^{a}\gamma^{\mu}\ell_{3L}\right)\left(V_{Lj} \blep[j] \sigma_{a}\gamma_{\mu}\lep[i]\vld[i] \right)~,
\end{eqnarray}
that leads to 
\begin{equation}
\frac{ \Gamma(\tau\to\mu\nu\overline{\nu}) }{   \Gamma(\tau\to e\nu\overline{\nu}) } = \left[
\frac{ \Gamma(\tau\to\mu\nu\overline{\nu})  }{ \Gamma(\tau\to e \nu\overline{\nu}) }  \right]_{\rm SM} \times 
\left| 1+2 C^{\ell-2}_{04} |V_L|^2 \right|^2~.
\label{eq:Gtaumue}
\ee
Using $\Gamma(\tau\to\mu\nu\overline{\nu})^{\rm exp}/\Gamma(\tau\to e\nu\overline{\nu})^{\rm exp}=0.9762 \pm 0.0028$
and $\Gamma(\tau\to\mu\nu\overline{\nu})_{\rm SM}/\Gamma(\tau\to e\nu\overline{\nu})^{\rm SM}= 0.9726$ we find 
\be
 \Re\left( C^{\ell-2}_{04} |V_L|^2 \right) = (0.95 \pm 0.70) \times 10^{-3}~,
\ee
 which is perfectly consistent with the power-counting expectation 
 $ \Re\left( C^{\ell-2}_{04} |V_L|^2 \right) =10^{-3}$ obtained for $\epslp \approx 0.1$.

\subsection{Upper bound on $|V_{L_1}/V_{L_2}|$}
\label{sect:Ve_max}

We conclude this Section with a na\"\i ve estimate 
of the maximal value of $|V_{L_1}/V_{L_2}|$ (or the electron component of the lepton spurion), which can  
regarded as a tuning in the lepton-flavor space of the EFT.  Assuming $|V_L|=\epslp =O(0.1)$, as required to explain the 
$\RK$ anomaly, the $|V_{L_1}/V_{L_2}|$  ratio is strongly bounded by $\mu \to e$  LFV processes. 
Employing the power-counting scheme defined in Section~\ref{sect:EFTexp}, the bounds dictated by the present experimental 
bounds on $\mu \to e$  conversion in Nuclei and $\cB(\mu \to 3e)$ turn out to be very similar. 
Focusing on the latter, the power-counting scheme implies 
\be
\cA(\mu \to 3e) \propto  (\epslp)^2 (\epsl{L})^2  \left(\frac{V_{L_1}}{V_{L_2}}\right)^2~.
\ee
Taking into account also the overall-suppression scale we get
\be
\cB(\mu \to 3e) \approx  10^{-8} \times \left( \frac{\epslp}{10^{-1}} \right)^4 
\left( \frac{\epsl{L}}{ 0.3 } \right)^4  \left|\frac{V_{L_1}}{V_{L_2}}\right|^2 < 1.0 \times 10^{-12}~,
\ee
where the last inequality corresponds to the present experimental constraint~\cite{Olive:2016xmw}.
As can be seen, for $|V_{L_1}/V_{L_2}| < 0.01$  the experimental bound
is satisfied. This ratio is significantly smaller than the corresponding 
$|V_{Q_d}/V_{Q_s}|$ ratio in the quark sector,
 but it is not unnatural given the observed hierarchies 
in the charged lepton mass matrix ($m_e/m_\mu \approx 5 \times 10^{-3}$).
 
\section{Conclusions}

In this paper we have analysed the consistency of the 
$R^{\tau/\ell}_{D^{(*)}}$ and $\RK$ 
anomalies 
with all available low-energy observables,
in the context of an EFT based on the $U(2)_q \times U(2)_\ell \times {\mathcal G}_{R}$ flavor symmetry defined in Eq.~(\ref{eq:flavsymm}).
The $R^{\tau/\ell}_{D^{(*)}}$ anomaly, if interpreted as a signal of NP, 
necessarily points toward a low effective scale for the EFT, slightly  below 1 TeV.
As a result, despite the MFV-like protection 
implied by the flavor symmetry, the latter  
is not enough to guarantee a natural consistency of the EFT with the  
tight constraints from various low-energy processes (most notably
precision measurements in $B$ and $\tau$ physics).
However, as we have shown,
a consistent picture for all low-energy observables 
can be obtained under  the additional dynamical assumption
that the NP sector is coupled preferentially 
to third generation SM fermions (or the singlets of the flavor symmetry). 

In the EFT context,  this dynamical  assumptions  can be realised	 
in general terms via the rescaling of fields (and operators) 
that we have identified in Sect.~\ref{sect:EFTexp}. This rescaling of the field 
leas to a modified power counting, and the resulting EFT turns out to be 
rather  coherent. 
Still some tuning of the EFT parameters are necessary in order to satisfy 
constraints from processes involving light quarks and leptons. 
More precisely, we have identified two main sources of tuning, both quantifiable around the $10\%$ level.
The first one is an alignment  in (quark) flavor space:  
the  flavor singlets need to be closely aligned to left-handed bottom quarks 
in order to satisfy the constraints from $B_{s(d)}$ mixing. The second one is a $O(10\%)$ cancellation 
of two independent terms in order to justify the absence of NP effects in $\cB(\tau \to \mu\nu\bar{\nu})$. 
Modulo these two tunings, the EFT allows us to accommodate  non-vanishing NP
contributions to $R^{\tau/\ell}_{D^{(*)}}$ and $\RK$  at the level of present anomalies, 
and contributions to the other observables below (or within) current uncertainties
for natural values of the other free parameters, as summarised in 
Table~\ref{tbl:results}.

The analysis of all existing bounds presented in Sect.~\ref{sect:Consistency}
can also be used to identify which are the most promising observables 
to obtain further evidences  of NP in this framework. In addition to the 
model-independent  confirmation of the anomalies in other $B$ decays 
(both charged and netural-current transitions),  the EFT construction has 
allowed us to identify there particularly interesting sets of 
observables in $\tau$ decays.
\begin{itemize}
\item[I.] \underline{LFV $\tau$ decays.} The branching ratios of  both purely leptonic
and semi-leptonic LFV $\tau$ decays can easily exceed the $10^{-9}$ level. 
\item[II.] \underline{Precision measurements of $\cB(\tau \to \ell \nu\bar\nu)$.}  
Violations of  $\mu/e$ universality and, more generally, deviations from the SM predictions in $\cB(\tau \to \ell \nu\bar\nu)$
are expected at the few per-mil level.
\item[III.] \underline{The determination of $|V_{us}|$ from $\tau$ decays.} 
Due to the breaking of  LFU, the $|V_{us}|$ determination from $\tau$ vs.~$K$ decays  
can differ at the $1\%$ level. 
\end{itemize}
While the first two categories have already been widely discussed in the literature (see e.g.
Ref.~\cite{Greljo:2015mma,Feruglio:2016gvd}), 
the last one has been identified for the first time by the present analysis.
In all these cases NP effects are expected just below current experimental sensitivities. 
Improved measurements of these observables could therefore provide a very valuable 
tool to provide further evidences or to falsify this framework in the near future.

\section*{Acknowledgements}
We thank Ferruccio Feruglio, Admir Greljo, Paride Paradisi, and Andrea Pattori,  for useful  discussions and
comments on the manuscript. This research was supported in
part by the Swiss National Science Foundation (SNF) under contract 200021-159720.

\appendix
\section{Hadronic Form Factors for $B\to V$ or $B\to P$ transitions}
\label{app:A}
We need to express explicitly the hadronic matrix elements through Lorentz invariant form factors. for $B\to P$ transitions, where $P$ is any pseudo-scalar meson, we have \cite{Becirevic:2016hea}:
\begin{align}
\label{pseudo_scalar_ff}
\langle P(k)\vert\overline{q}_{i}\gamma_{\mu}b\vert\overline{B}(p)\rangle=&\left[(p+k)_{\mu}-\frac{m_{B}^2-m_{P}^2}{q^2}q_{\mu}\right]f_{+}(q^2)+q_{\mu}\frac{m_{B}^2-m_{P}^2}{q^2}f_{0}(q^2) \\
\langle P(k)\vert[\overline{q}_{i}b](\mu)\vert\overline{B}(p)\rangle=&\frac{1}{m_{b}(\mu)-m_{q_{i}}(\mu)}q^{\mu}\langle P(k)\vert\overline{q}_{i}\gamma_{\mu}b\vert\overline{B}(p)\rangle=\frac{m_{B}^2-m_{P}^2}{m_{b}(\mu)-m_{q_{i}}(\mu)}f_{0}(q^2).
\end{align}
Instead, for $B\to V$ transitions, where V is a vector meson, we use:
\begin{align}
\label{vector_ff}
\langle V(k,\eta)\vert\overline{q}_{i}\gamma_{\mu}b\vert\overline{B}(p)\rangle=& \ i\epsilon_{\mu\nu\rho\sigma}\eta^{\nu*}p^{\rho}k^{\sigma}\frac{2V(q^2)}{m_{B}+m_{V}}, \\
\langle V(k,\eta)\vert\overline{q}_{i}\gamma_{\mu}\gamma_{5}b\vert\overline{B}(p)\rangle=& \ \eta^{*}_{\mu}(m_{B}+m_{V})A_{1}(q^2)-(p+k)_{\mu}(\eta^{*}\cdot q)\frac{A_{2}(q^2)}{m_{B}+m_{V}} \notag\\
&-q_{\mu}(\eta^{*}\cdot q)\frac{2m_{V}}{q^2}[A_{3}(q^2)-A_{0}(q^2)], \\
\langle V(k,\eta)\vert[\overline{q}_{i}\gamma_{5}b](\mu)\vert\overline{B}(p)\rangle=& \ -\frac{1}{m_{b}(\mu)+m_{q_{i}}(\mu)}q_{\mu}\langle V(k,\eta)\vert\overline{q}_{i}\gamma_{\mu}\gamma_{5}b\vert\overline{B}(p)\rangle \notag\\	
=& \ (\eta^{*}\cdot q)\frac{2 m_{V}}{m_{b}(\mu)+m_{q_{i}}(\mu)}A_{0}(q^2).
\end{align}
where we can express $A_{3}(q^2)$ as:
\begin{equation}
A_{3}(q^2)=\frac{m_{B}+m_{V}}{2 m_{V}}A_{1}(q^2)-\frac{m_{B}-m_{V}}{2 m_{V}}A_{2}(q^2),
\end{equation}
and we changed the form factors basis in
\begin{align}
V(q^2)= \ &\frac{m_{B}(m_{B}+m_{V})}{\sqrt{2\lambda}} F_{\perp} , \\
A_{1}(q^2)= \ &\frac{m_{B}}{\sqrt{2}(m_{B}+m_{V})}F_{\parallel}  , \\
A_{2}(q^2)= \ &  -\frac{2m_{V}m_{B}^2(m_{B}+m_{V})}{\lambda}F_{0}(q^2)+\frac{m_{B}(m_{B}+m_{V})(m_{B}^2-m_{V}^2-q^2)}{\sqrt{2}\lambda}A_{1}(q^2), \\
A_{0}(q^2)= \ &\frac{m_{B}^2}{\sqrt{\lambda}}F_{t}(q^2).
\end{align}

\section{Differential decay width for $B\to K \ell\overline{\ell}$}
\label{app:B}
In this Appendix we intend to give the complete expression for the differential decay width of the process  $B\to K\ell\overline{\ell}$, where $\ell=\mu,\tau,\nu$. For this purpose we keep the full dependence from the lepton mass, which gives a non negligible contribution in the case $\ell=\tau$.  \\
The most general Lagrangian that arises from the operators in \Tables{table:sl_1}{table:sl_2} assumes the form:
\begin{equation}
\mathcal{L}(b\rightarrow s\ell\overline{\ell})=-\frac{2G_{F}}{\sqrt{2}}\frac{\alpha_{e}}{2\pi}V_{ts}^* V_{tb}\left[C_{9}\mathcal{O}_{9}+C_{10}\mathcal{O}_{10} +C_{S}(\mathcal{O}_{S}-\mathcal{O}_{P})\right] \, ,
\end{equation}
where the operators are defined as
\begin{equation}
\label{FCNC_wilson coefficients}
\begin{aligned}
\mathcal{O}_{9}=&(\overline{s}\gamma_{\mu}P_{L}b)(\overline{\ell}\gamma_{\mu}\ell) & \mathcal{O}_{10}=&(\overline{s}\gamma_{\mu}P_{L})b(\overline{\ell}\gamma_{\mu}\gamma_{5}\ell)  \, ,\\
\mathcal{O}_S=&(\overline{s}P_{R}b)(\ell\overline{\ell})  & \mathcal{O}_P=&(\overline{s}P_{R}b)(\overline{\ell}\gamma_{5}\ell) \, .
\end{aligned}
\end{equation}
By mean of explicit calculation, we can write the double differential decay width as
\begin{equation}
\frac{\mathrm{d}^{2}\Gamma}{\mathrm{d}\cos\theta\mathrm{d}q^2}=a_{\ell}+b_{\ell}\cos\theta+c_{\ell}\cos^{2}\theta
\end{equation}
and the coefficients are
\begin{align}
\frac{4 a_{\ell}}{\Gamma_{0}}=& (\vert C_{9}\vert^2+\vert C_{10}\vert^2)f_{+}^2\lambda+4\vert C_{10}\vert^2\frac{m_{\ell}^2}{q^2}\left[f_{0}^2(m_{B}^2-m_{K}^2)^2-f_{+}^2\lambda\right] \notag\\
-&\frac{4C_{S}C_{10}f_{0}^2m_{\ell}(m_{B}^2-m_{K}^2)^2}{m_{b}-m_{s}}+\frac{2C_{S}^2 f_{0}^2(m_{B}^2-m_{K}^2)^2(q^2-2m_{\ell}^2	)}{(m_{b}-m_{s})^2} \, , \\
\frac{b_{\ell}}{\Gamma_{0}}=&  \ \frac{C_{S}C_{9} f_{+}f_{0} \sqrt{\lambda}m_{\ell}(m_{B}^2-m_{K}^2)\beta_{\ell}}{m_{b}-m_{s}}\, , \\
\frac{4 c_{\ell}}{\Gamma_{0}}=& \ \beta_{\ell}^{2}\lambda f_{+}^{2}\left(\vert C_9\vert^2+\vert C_{10}\vert^2\right) \, . 
\end{align}
where
\begin{align}
\beta_{\ell}=& \sqrt{1-\frac{4m_{\ell}^2}{q^2}},  &  \Gamma_{0}&=\frac{\alpha_{e}^2 G_{F}^2 \sqrt{\lambda } \beta_{\ell} \vert V_{tb}V^{*}_{ts}\vert^2}{512 \  \pi ^5 m_{B}^3} ,
\end{align}
\begin{equation}
 \lambda= m_{B}^4+m_{K}^4+q^4-2m_{B}^2 q^2-2m_{K}^2 q^2-2m_{B}^2 m_{K}^2.
\end{equation}
Performing the angular integration we get the following differential decay width:
\begin{align}
\label{width_BtoKtautau}
\frac{\text{d}\Gamma}{\text{d}q^2}(B\to K\ell\overline{\ell})=&\frac{\Gamma_{0}}{6}\bigg[(\vert C_{9}\vert^2+\vert C_{10}\vert^2)(3-\beta_{\ell}^2)f_{+}^2\lambda+12	\vert C_{10}\vert^2\frac{m_{\ell}^2}{q^2}\left(f_{0}^2(m_{B}^2-m_{K}^2)^2-f_{+}^2\lambda\right)\bigg]\notag \\
-&2\Gamma_{0}C_{S}C_{10}\frac{f_{0}^2
   m_{\ell}
   \left({m_B}^2-m_{K}^2\right)^2 }{m_{b}-m_{s}}+\Gamma_{0}C_{S}^2\frac{f_{0}^2   \left(m_{B}^2-m_{K}^2\right)^2 \left(q^2-2 m_{\ell}^2\right)}{(m_{b}-m_{s})^2}
\end{align}

\bibliographystyle{my.bst}

{\footnotesize
\bibliography{paper}

\providecommand{\href}[2]{#2}\begingroup\raggedright\begin{thebibliography}{10}

\bibitem{Amhis:2016xyh}
Y.~Amhis {\em et al.,}
\href{http://arxiv.org/abs/1612.07233}{{\tt arXiv:1612.07233 [hep-ex]}}.

\bibitem{Lees:2013uzd}
{\bf BaBar}, J.~P. Lees {\em et al.,}
  \href{http://dx.doi.org/10.1103/PhysRevD.88.072012}{{\em Phys. Rev.} {\bf
  D88} (2013) no.~7, 072012},
\href{http://arxiv.org/abs/1303.0571}{{\tt arXiv:1303.0571 [hep-ex]}}.

\bibitem{Hirose:2016wfn}
{\bf Belle}, S.~Hirose {\em et al.,}
\href{http://arxiv.org/abs/1612.00529}{{\tt arXiv:1612.00529 [hep-ex]}}.

\bibitem{Aaij:2015yra}
{\bf LHCb}, R.~Aaij {\em et al.,}
  \href{http://dx.doi.org/10.1103/PhysRevLett.115.159901,
  10.1103/PhysRevLett.115.111803}{{\em Phys. Rev. Lett.} {\bf 115} (2015)
  no.~11, 111803}, \href{http://arxiv.org/abs/1506.08614}{{\tt arXiv:1506.08614
  [hep-ex]}}.
[Addendum: Phys. Rev. Lett.115,no.15,159901(2015)].

\bibitem{Fajfer:2012vx}
S.~Fajfer, J.~F. Kamenik, and I.~Nisandzic,
  \href{http://dx.doi.org/10.1103/PhysRevD.85.094025}{{\em Phys. Rev.} {\bf
  D85} (2012)  094025},
\href{http://arxiv.org/abs/1203.2654}{{\tt arXiv:1203.2654 [hep-ph]}}.

\bibitem{Aoki:2016frl}
S.~Aoki {\em et al.,}
\href{http://arxiv.org/abs/1607.00299}{{\tt arXiv:1607.00299 [hep-lat]}}.

\bibitem{Aaij:2014ora}
{\bf LHCb}, R.~Aaij {\em et al.,}
  \href{http://dx.doi.org/10.1103/PhysRevLett.113.151601}{{\em Phys. Rev.
  Lett.} {\bf 113} (2014)  151601},
\href{http://arxiv.org/abs/1406.6482}{{\tt arXiv:1406.6482 [hep-ex]}}.

\bibitem{Bordone:2016gaq}
M.~Bordone, G.~Isidori, and A.~Pattori,
  \href{http://dx.doi.org/10.1140/epjc/s10052-016-4274-7}{{\em Eur. Phys. J.}
  {\bf C76} (2016) no.~8, 440},
\href{http://arxiv.org/abs/1605.07633}{{\tt arXiv:1605.07633 [hep-ph]}}.

\bibitem{Aaij:2015oid}
{\bf LHCb}, R.~Aaij {\em et al.,}
  \href{http://dx.doi.org/10.1007/JHEP02(2016)104}{{\em JHEP} {\bf 02} (2016)
  104},
\href{http://arxiv.org/abs/1512.04442}{{\tt arXiv:1512.04442 [hep-ex]}}.

\bibitem{Wehle:2016yoi}
{\bf Belle}, S.~Wehle {\em et al.,}
\href{http://arxiv.org/abs/1612.05014}{{\tt arXiv:1612.05014 [hep-ex]}}.

\bibitem{Ciuchini:2015qxb}
M.~Ciuchini, M.~Fedele, E.~Franco, S.~Mishima, A.~Paul, L.~Silvestrini, and
  M.~Valli, \href{http://dx.doi.org/10.1007/JHEP06(2016)116}{{\em JHEP} {\bf
  06} (2016)  116},
\href{http://arxiv.org/abs/1512.07157}{{\tt arXiv:1512.07157 [hep-ph]}}.

\bibitem{Descotes-Genon:2015uva}
S.~Descotes-Genon, L.~Hofer, J.~Matias, and J.~Virto,
  \href{http://dx.doi.org/10.1007/JHEP06(2016)092}{{\em JHEP} {\bf 06} (2016)
  092},
\href{http://arxiv.org/abs/1510.04239}{{\tt arXiv:1510.04239 [hep-ph]}}.

\bibitem{Altmannshofer:2015sma}
W.~Altmannshofer and D.~M. Straub, in {\em {Proceedings, 50th Rencontres de
  Moriond Electroweak Interactions and Unified Theories (La Thuile, Italy,
  March 14-21, 2015)}}.
\newblock
\href{http://arxiv.org/abs/1503.06199}{{\tt arXiv:1503.06199 [hep-ph]}}.
\newblock

\bibitem{Hurth:2016fbr}
T.~Hurth, F.~Mahmoudi, and S.~Neshatpour,
  \href{http://dx.doi.org/10.1016/j.nuclphysb.2016.05.022}{{\em Nucl. Phys.}
  {\bf B909} (2016)  737--777},
\href{http://arxiv.org/abs/1603.00865}{{\tt arXiv:1603.00865 [hep-ph]}}.

\bibitem{Bhattacharya:2014wla}
B.~Bhattacharya, A.~Datta, D.~London, and S.~Shivashankara,
  \href{http://dx.doi.org/10.1016/j.physletb.2015.02.011}{{\em Phys. Lett.}
  {\bf B742} (2015)  370--374},
\href{http://arxiv.org/abs/1412.7164}{{\tt arXiv:1412.7164 [hep-ph]}}.

\bibitem{Alonso:2015sja}
R.~Alonso, B.~Grinstein, and J.~Martin~Camalich,
  \href{http://dx.doi.org/10.1007/JHEP10(2015)184}{{\em JHEP} {\bf 10} (2015)
  184},
\href{http://arxiv.org/abs/1505.05164}{{\tt arXiv:1505.05164 [hep-ph]}}.

\bibitem{Greljo:2015mma}
A.~Greljo, G.~Isidori, and D.~Marzocca,
  \href{http://dx.doi.org/10.1007/JHEP07(2015)142}{{\em JHEP} {\bf 07} (2015)
  142},
\href{http://arxiv.org/abs/1506.01705}{{\tt arXiv:1506.01705 [hep-ph]}}.

\bibitem{Calibbi:2015kma}
L.~Calibbi, A.~Crivellin, and T.~Ota,
  \href{http://dx.doi.org/10.1103/PhysRevLett.115.181801}{{\em Phys. Rev.
  Lett.} {\bf 115} (2015)  181801},
\href{http://arxiv.org/abs/1506.02661}{{\tt arXiv:1506.02661 [hep-ph]}}.

\bibitem{Bauer:2015knc}
M.~Bauer and M.~Neubert,
  \href{http://dx.doi.org/10.1103/PhysRevLett.116.141802}{{\em Phys. Rev.
  Lett.} {\bf 116} (2016) no.~14, 141802},
\href{http://arxiv.org/abs/1511.01900}{{\tt arXiv:1511.01900 [hep-ph]}}.

\bibitem{Fajfer:2015ycq}
S.~Fajfer and N.~Ko?nik,
  \href{http://dx.doi.org/10.1016/j.physletb.2016.02.018}{{\em Phys. Lett.}
  {\bf B755} (2016)  270--274},
\href{http://arxiv.org/abs/1511.06024}{{\tt arXiv:1511.06024 [hep-ph]}}.

\bibitem{Barbieri:2015yvd}
R.~Barbieri, G.~Isidori, A.~Pattori, and F.~Senia,
  \href{http://dx.doi.org/10.1140/epjc/s10052-016-3905-3}{{\em Eur. Phys. J.}
  {\bf C76} (2016) no.~2, 67},
\href{http://arxiv.org/abs/1512.01560}{{\tt arXiv:1512.01560 [hep-ph]}}.

\bibitem{Das:2016vkr}
D.~Das, C.~Hati, G.~Kumar, and N.~Mahajan,
  \href{http://dx.doi.org/10.1103/PhysRevD.94.055034}{{\em Phys. Rev.} {\bf
  D94} (2016)  055034},
\href{http://arxiv.org/abs/1605.06313}{{\tt arXiv:1605.06313 [hep-ph]}}.

\bibitem{Boucenna:2016qad}
S.~M. Boucenna, A.~Celis, J.~Fuentes-Martin, A.~Vicente, and J.~Virto,
  \href{http://dx.doi.org/10.1007/JHEP12(2016)059}{{\em JHEP} {\bf 12} (2016)
  059},
\href{http://arxiv.org/abs/1608.01349}{{\tt arXiv:1608.01349 [hep-ph]}}.

\bibitem{Becirevic:2016yqi}
D.~Becirevic, S.~Fajfer, N.~Kosnik, and O.~Sumensari,
  \href{http://dx.doi.org/10.1103/PhysRevD.94.115021}{{\em Phys. Rev.} {\bf
  D94} (2016) no.~11, 115021},
\href{http://arxiv.org/abs/1608.08501}{{\tt arXiv:1608.08501 [hep-ph]}}.

\bibitem{Hiller:2016kry}
G.~Hiller, D.~Loose, and K.~Schoenwald,
  \href{http://dx.doi.org/10.1007/JHEP12(2016)027}{{\em JHEP} {\bf 12} (2016)
  027},
\href{http://arxiv.org/abs/1609.08895}{{\tt arXiv:1609.08895 [hep-ph]}}.

\bibitem{Bhattacharya:2016mcc}
B.~Bhattacharya, A.~Datta, J.-P. Gu\'evin, D.~London, and R.~Watanabe,
  \href{http://dx.doi.org/10.1007/JHEP01(2017)015}{{\em JHEP} {\bf 01} (2017)
  015},
\href{http://arxiv.org/abs/1609.09078}{{\tt arXiv:1609.09078 [hep-ph]}}.

\bibitem{Barbieri:2011ci}
R.~Barbieri, G.~Isidori, J.~Jones-Perez, P.~Lodone, and D.~M. Straub,
  \href{http://dx.doi.org/10.1140/epjc/s10052-011-1725-z}{{\em Eur. Phys. J.}
  {\bf C71} (2011)  1725},
\href{http://arxiv.org/abs/1105.2296}{{\tt arXiv:1105.2296 [hep-ph]}}.

\bibitem{Barbieri:2012uh}
R.~Barbieri, D.~Buttazzo, F.~Sala, and D.~M. Straub,
  \href{http://dx.doi.org/10.1007/JHEP07(2012)181}{{\em JHEP} {\bf 07} (2012)
  181},
\href{http://arxiv.org/abs/1203.4218}{{\tt arXiv:1203.4218 [hep-ph]}}.

\bibitem{Buttazzo:2016kid}
D.~Buttazzo, A.~Greljo, G.~Isidori, and D.~Marzocca,
  \href{http://dx.doi.org/10.1007/JHEP08(2016)035}{{\em JHEP} {\bf 08} (2016)
  035},
\href{http://arxiv.org/abs/1604.03940}{{\tt arXiv:1604.03940 [hep-ph]}}.

\bibitem{Barbieri:2016las}
R.~Barbieri, C.~W. Murphy, and F.~Senia,
  \href{http://dx.doi.org/10.1140/epjc/s10052-016-4578-7}{{\em Eur. Phys. J.}
  {\bf C77} (2017) no.~1, 8},
\href{http://arxiv.org/abs/1611.04930}{{\tt arXiv:1611.04930 [hep-ph]}}.

\bibitem{Faroughy:2016osc}
D.~A. Faroughy, A.~Greljo, and J.~F. Kamenik,
  \href{http://dx.doi.org/10.1016/j.physletb.2016.11.011}{{\em Phys. Lett.}
  {\bf B764} (2017)  126--134},
\href{http://arxiv.org/abs/1609.07138}{{\tt arXiv:1609.07138 [hep-ph]}}.

\bibitem{Feruglio:2016gvd}
F.~Feruglio, P.~Paradisi, and A.~Pattori,
\href{http://arxiv.org/abs/1606.00524}{{\tt arXiv:1606.00524 [hep-ph]}}.

\bibitem{DAmbrosio:2002vsn}
G.~D'Ambrosio, G.~F. Giudice, G.~Isidori, and A.~Strumia,
  \href{http://dx.doi.org/10.1016/S0550-3213(02)00836-2}{{\em Nucl. Phys.} {\bf
  B645} (2002)  155--187},
\href{http://arxiv.org/abs/hep-ph/0207036}{{\tt arXiv:hep-ph/0207036
  [hep-ph]}}.

\bibitem{EOS}
D.~van Dyk {\em et al.,}, ``{EOS - A HEP Program for Flavour Observables}.''
\newblock \url{http://github.com/eos/eos}.

\bibitem{Olive:2016xmw}
{\bf Particle Data Group}, C.~Patrignani {\em et al.,}
\href{http://dx.doi.org/10.1088/1674-1137/40/10/100001}{{\em Chin. Phys.} {\bf
  C40} (2016) no.~10, 100001}.

\bibitem{Becirevic:2016hea}
D.~Becirevic, S.~Fajfer, I.~Nisandzic, and A.~Tayduganov,
\href{http://arxiv.org/abs/1602.03030}{{\tt arXiv:1602.03030 [hep-ph]}}.

\bibitem{Pich:2013lsa}
A.~Pich \href{http://dx.doi.org/10.1016/j.ppnp.2013.11.002}{{\em Prog. Part.
  Nucl. Phys.} {\bf 75} (2014)  41--85},
\href{http://arxiv.org/abs/1310.7922}{{\tt arXiv:1310.7922 [hep-ph]}}.

\bibitem{Buchalla:1995vs}
G.~Buchalla, A.~J. Buras, and M.~E. Lautenbacher,
  \href{http://dx.doi.org/10.1103/RevModPhys.68.1125}{{\em Rev. Mod. Phys.}
  {\bf 68} (1996)  1125--1144},
\href{http://arxiv.org/abs/hep-ph/9512380}{{\tt arXiv:hep-ph/9512380
  [hep-ph]}}.

\bibitem{Barbieri:2014tja}
R.~Barbieri, D.~Buttazzo, F.~Sala, and D.~M. Straub,
  \href{http://dx.doi.org/10.1007/JHEP05(2014)105}{{\em JHEP} {\bf 05} (2014)
  105},
\href{http://arxiv.org/abs/1402.6677}{{\tt arXiv:1402.6677 [hep-ph]}}.

\bibitem{TheBaBar:2016xwe}
{\bf BaBar}
\href{http://arxiv.org/abs/1605.09637}{{\tt arXiv:1605.09637 [hep-ex]}}.

\bibitem{Brod:2010hi}
J.~Brod, M.~Gorbahn, and E.~Stamou,
  \href{http://dx.doi.org/10.1103/PhysRevD.83.034030}{{\em Phys. Rev.} {\bf
  D83} (2011)  034030},
\href{http://arxiv.org/abs/1009.0947}{{\tt arXiv:1009.0947 [hep-ph]}}.

\bibitem{Bouchard:2013pna}
{\bf HPQCD}, C.~Bouchard, G.~P. Lepage, C.~Monahan, H.~Na, and J.~Shigemitsu,
  \href{http://dx.doi.org/10.1103/PhysRevD.88.079901,
  10.1103/PhysRevD.88.054509}{{\em Phys. Rev.} {\bf D88} (2013) no.~5, 054509},
  \href{http://arxiv.org/abs/1306.2384}{{\tt arXiv:1306.2384 [hep-lat]}}.
[Erratum: Phys. Rev.D88,no.7,079901(2013)].

\bibitem{Gelhausen:2013wia}
P.~Gelhausen, A.~Khodjamirian, A.~A. Pivovarov, and D.~Rosenthal,
  \href{http://dx.doi.org/10.1103/PhysRevD.88.014015,
  10.1103/PhysRevD.91.099901, 10.1103/PhysRevD.89.099901}{{\em Phys. Rev.} {\bf
  D88} (2013)  014015}, \href{http://arxiv.org/abs/1305.5432}{{\tt
  arXiv:1305.5432 [hep-ph]}}.
[Erratum: Phys. Rev.D91,099901(2015)].

\bibitem{Alte:2016yuw}
S.~Alte, M.~K{\"o}nig, and M.~Neubert,
  \href{http://dx.doi.org/10.1007/JHEP12(2016)037}{{\em JHEP} {\bf 12} (2016)
  037},
\href{http://arxiv.org/abs/1609.06310}{{\tt arXiv:1609.06310 [hep-ph]}}.

\bibitem{Glashow:2014iga}
S.~L. Glashow, D.~Guadagnoli, and K.~Lane,
  \href{http://dx.doi.org/10.1103/PhysRevLett.114.091801}{{\em Phys. Rev.
  Lett.} {\bf 114} (2015)  091801},
\href{http://arxiv.org/abs/1411.0565}{{\tt arXiv:1411.0565 [hep-ph]}}.

\end{thebibliography}\endgroup
}

\end{document}